\documentclass[referee]{raa}            % referee version: for submission

\usepackage{graphicx,times,subfigure}             %for PS/EPS graphics inclusion, new
\usepackage{float}                                 %user added

\begin{document}

   \title{A Study on Universal Observation Control System And Its Application For LAMOST
%\footnotetext{$*$ Supported by the National Natural Science Foundation of China.}
}
%   \subtitle{I. Place Your Subtitle Here}

   \volnopage{Vol.0 (200x) No.0, 000--000}      %%preserved for Editor. DOn't remove!
   \setcounter{page}{1}          %%starting page, preserved for Editor. DOn't remove!

   \author{Zheng Wang
      \inst{1,2}
   \and Yuan Tian
      \inst{1,2}
   \and Jian Li
      \inst{1}
   \and Zi-Huang Cao
      \inst{1}  
   \and Yong-Heng Zhao
      \inst{1,2}
   }

   \institute{Key Laboratory of Optical Astronomy, National Astronomical Observatories, Chinese Academy of Sciences,
             Beijing 100101, China; {\it wzheng@bao.ac.cn}\\
        \and
             University of Chinese Academy of Sciences, Beijing 100049, China
   }

   \date{Received~~2009 month day; accepted~~2009~~month day}

\abstract{The observatory control system (OCS), a part of automated control of Large Sky Area Multi-Object Fibre Spectroscopic Telescope (LAMOST), runs on the CentOS6 platform and implements the communication between modules based on Common Object Request Broker Architecture (CORBA). However, CORBA is complicated and has limited development support; moreover, the official support for CentOS6 has ended. OCS inherently has some shortcomings such as the over-concentration of control and the blocking of device status processing, which hinder the realization of automated observation control of LAMOST. Therefore, this study designs and implements a universal observation control system (UOCS) for optical telescopes. The UOCS takes the device command as the basic execution unit, controls the device execution logic using observation script, controls the observation logic by event-driven messaging, and realizes mutual decoupling between modules via a distributed control mode, thereby ensuring high system robustness. The UOCS performs significantly better than OCS in terms of the observation performance, operator complexity, and communication error. Currently, UOCS is applied to the automated control of some devices and subsystems in LAMOST observation. It will be applied to the automated observation control of Multi-channel Photometric Survey Telescope by 2021.
}

   \authorrunning{Z.Wang  et al.}            %author_head in even pages
   \titlerunning{A Study on Universal Observation Control System And Its Application For LAMOST}  % title_head in odd pages

   \maketitle
%% The author head (on even pages) and the title head (on odd pages) will be
%% automatically extracted from \author{} and \title{}. Whenever the title is too long,
%% you will be asked to supply a shorter one by inserting either \authorrunning{} or
%% \titlerunning{} before \maketitle. Anyway, you can specify your own heads.
%%
%%
%% Note: In the following text body of your manuscript, please note several differences from
%%       other major journals:
%% (1) \subsection{Please Capitalize the First Letter of Each Notional Word in Subsection Title}
%% (2) Please Capitalize the First Letter of Each Notional Word in all tables' captions

%
%________________________________________________ sections below
%
\section{Introduction}           %% first-level sections will be auto-capitalized
\label{sect:intro}

The Large Sky Area Multi-Object Fibre Spectroscopic Telescope (LAMOST), also known as the Guo Shoujing Telescope, is a Schmidt optical reflecting telescope(~\cite{cui2012}), located in the Xinglong Observatory of the National Astronomical Observatory of China (NAOC). Owing to its large aperture and wide field-of-view, LAMOST can observe up to 4000 objects simultaneously in a single exposure, making it the most efficient telescope in terms of spectral acquisition worldwide(~\cite{zhao2015}). To date, LAMOST has obtained a catalogue of more than 10 million astronomical spectra\footnote{http://dr7.lamost.org/}.

An observatory control system (OCS), which is the communication center of a telescope, is a software for autonomous observatory control that integrates important functions, such as selecting the observation target, making observation flows, distributing commands, handling exceptions, controlling devices or subsystems, and human–computer interaction(~\cite{zhao2000}). Therefore, OCS is a key element in the autonomous control of a telescope, and it is pivotal in integrating the functions of the entire telescope control system.

The OCS is a distributed control system based on the CentOS6 computing platform, and it uses the Common Object Request Broker Architecture (CORBA) as the middleware(~\cite{wang2006}). The design and development of the OCS commenced in 2000 and was completed and applied to the LAMOST in 2008(~\cite{sun2008}). Since then, OCS has been providing vital support for observations using the LAMOST. However, as the LAMOST survey continues to gradually advance, a few shortcomings of the OCS have been revealed, some of which may even make the OCS defunct. Some of these shortcomings are listed below.

1) Difficulties in upgrading and maintenance. Updates and maintenance are no longer available since version 3.3 of CORBA (2012)\footnote{https://www.corba.org/}, making CORBA incompatible with newer operating systems, such as CentOS7. The ACE ORB (TAO), which forms the basis for the communication functions of the OCS, is a version of CORBA implemented based on the Adaptive Communication Environment (ACE) library. However, the TAO has only limited third-party development support. Moreover, because the official support for CentOS6 has ended, obtaining a stable and reliable system is difficult, rendering the security of OCS vulnerable.

2) Over-centralization of command control. Although the OCS adopts distributed control, it is distributed only among the internal modules. Furthermore, the device agent performs only the simple function of translating the control commands and states between the OCS and devices, and it plays no role in the business logic of control. The command executor, being the core component, integrates the tasks of sending commands, execution, and exception handling. This results in increased complexity of the business logic of internal control, a higher degree of coupling with other modules, and lower stability of the entire system. For example, if an error occurs in a parallel command, all parallel commands need to be re-executed, failing which no subsequent command could be executed.

3) Foreground and background are not separated. The necessity of separately restarting a single module was not fully considered in the design of OCS. Therefore, if an error arises, it can be rectified only by restarting the entire system—this interrupts the entire observation process, reducing the observation efficiency. 

4) Status transmission channel is indistinguishable. The OCS status is classified into command execution status and device normal status: the former is fed back to the command executor to control the observation process, while the latter is used to display the status of the subsystems or devices. No status is prioritized, and all statuses are transmitted via a single channel. Consequently, if the channel is blocked by the device’s normal status, the command executor cannot receive the command execution status, whose logic priority is higher than that of the device normal status. Thus, subsequent commands cannot be executed, resulting in interrupted observation flow. 

Owing to these shortcomings, the OCS can only be used for the manual control of some subsystems; however, the control of other subsystems also requires manual intervention through the subsystem of the control computer. This results in low observation efficiency and difficulties in operation and maintenance. Further, as mentioned above, OCS could become unusable because of the incompatibility of CORBA and newer operating systems (e.g., CentOS7). In view of these problems, a new control system must be developed and applied to the observation control of LAMOST.

Accordingly, this study designs a universal observation control system (UOCS) based on the asynchronous coroutine function of Python and implements it in an optical telescope. For a complete inclusion of the design concepts of OCS, the advantages and disadvantages of OCS are summarized and analyzed, and the overall reorganization design is performed comprehensively.Further, the developed UOCS is more advanced than OCS, and has improved suitability for diverse astronomical observation scenarios. Based on the device control command and the lightweight messaging middleware, ZeroMQ(~\cite{dworak2010}), an independent data and message bus and a device control system are developed for UOCS. The execution logic between the devices is controlled by observation scripts, and the serial–parallel distributed observation control is realized by combining the event and device commands. The UOCS performs command and exception handling using independent device and service modules, and it realizes human–computer interaction for the telescope by using the local user interface or network remote interface. The UOCS performs significantly better than OCS in terms of the observation performance, operator complexity, and communication error.

%% Authors can give a citation as 'Michel et al. 1992'.
%% You may also use \cite, \citep and \citet for citation, and use Table~1 or Figure~1
%% and so forth. Using \ref and \label for cross-references of Tables/Figures
%% is a good way in adjusting/adding/removing text, tables or figures.

\section{Design of the UOCS architecture}
\label{sect:DUA}
To reduce the degree of coupling of the internal modules and ensure a stable and efficient operation of the telescope, we adopted a distributed control structure with separated foreground and background for the UOCS (Figure 1).

\begin{figure}[!htbp]
\centering
\includegraphics[width=10cm, angle=0]{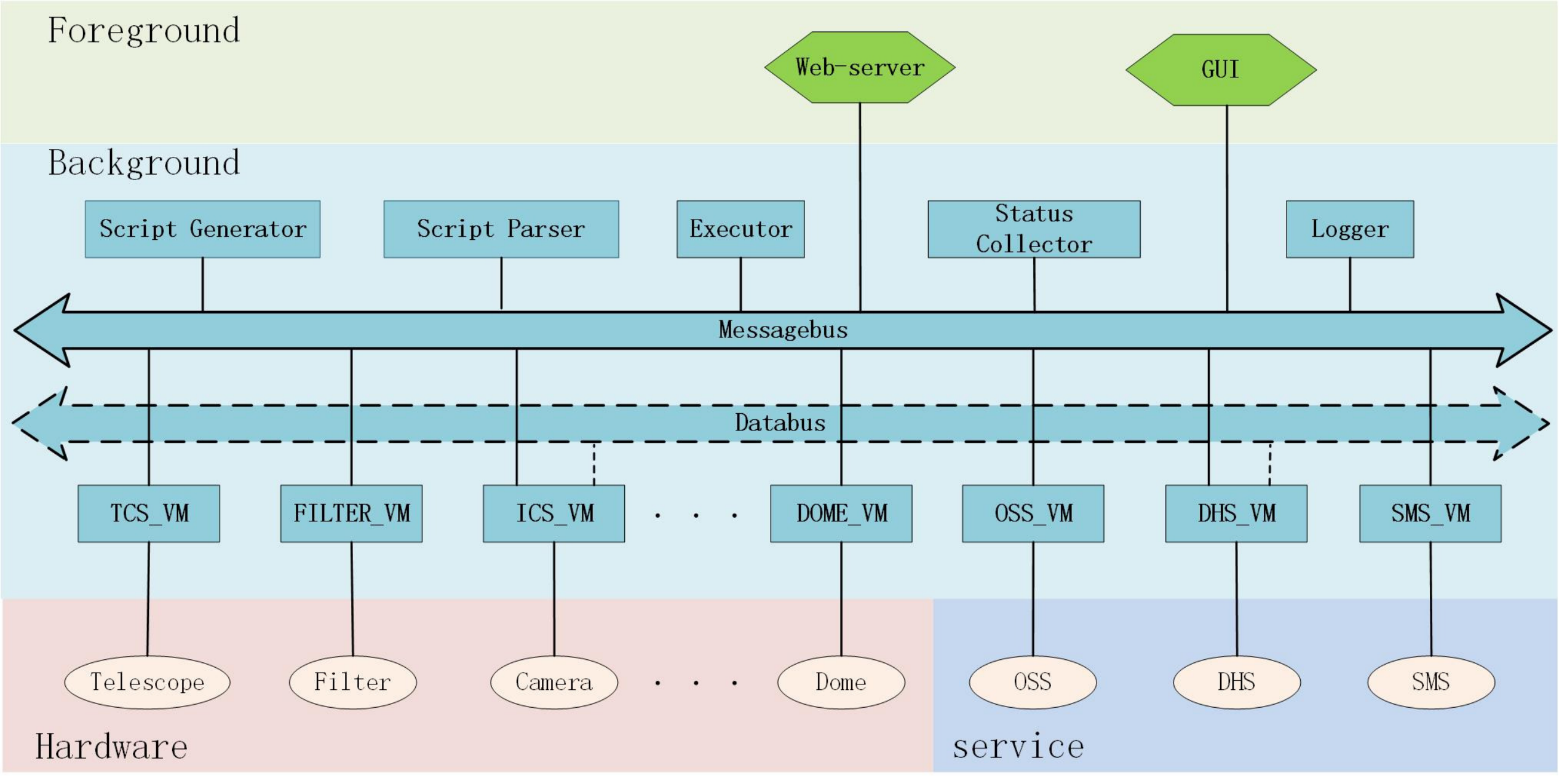}
\caption{Framework of the universal observation control system (UOCS)(GUI:Graphical User Interface;VM:Virtual Machine;TCS:Telescope Control System;ICS:Instrument Control System;OSS:Observation Schedule System;DHS:Data Handling System;SMS:Site Monitor System )}
\label{Fig1}
\end{figure}

The foreground handles the interaction between the telescope and the observer—it shows detailed information of each module of the telescope and generates observation control events and control commands for the devices and services. Based on the control mode, it can be divided into local control and remote web control. 

The background is the business logic center of the telescope observation control system, and it comprises the following three modules. (a) Communication bus module: it transmits the control commands, status information, and observation data between the modules. (b) Device module: a device directly controlled by the UOCS during observation; it can be divided into various types, such as camera module, filter module, optical system module, dome module, and the module for mounting telescope control (TCS). (c) Service module: it provides the algorithms, logic control, and other services for the UOCS. According to the type of service, service modules are classified into various types, including the observation target selection module, script generation module, script parse module, command execution module, status collection module, and site service module.

\subsection{Communication protocol }
The UOCS communication protocol can be divided into internal communication protocol and device communication protocol. The device communication protocol varies for each device and is not discussed here. In the initial stages of designing the internal communication protocol of the UOCS, we considered referring to the communication structure of RTS2 and using the classic TCP/IP protocol(~\cite{peter2006}, ~\cite{peter2008}). However, the complexity of the network structure worsens with the number of modules, which poses challenges to the development at the initial stages and maintenance at the later stages. Thus, we used the XPUB/XSUB pattern of ZeroMQ to connect all modules of UOCS. Each module generates messages that represent its own status and sends them to the communication bus, and at the same time, subscribes to the messages it needs to process from the communication bus.

According to the data type and the required effectiveness, the communication bus can be divided into two types: (a) message bus, which is used to transmit small amounts of data with high real-time performance, such as control commands and device status; and (b) data bus, which is used to transmit large amounts of data with low real-time performance, such as images and other data.

Messages comprise a SenderID and message body, and it can be divided into three types depending on the content of the message body: (a) command message, which is generated by the command executor and user interface to control devices; (b) status message, which includes the status of command execution and device information; and (c) event message, which is a real-time message generated by each module according to the running condition—it is used to perform logical control of the modules and has a unidirectional control flow.

\subsection{Script generator and parser}
Depending on the scientific objectives and observation procedures, the UOCS executes the control of multiple devices and service modules in serial, parallel, or serial–parallel combinations. Even for the same module composition, the observation procedures will vary for different telescopes with different scientific objectives. An observation script is a configuration file that describes the execution sequence of each device of the telescope; each observation task has a unique requirement, and all such requirements are realized by configuring different observation script files. The details on the UOCS observation task are discussed later in Section 4.2.2.

\begin{figure}[!htbp]
\centering
\includegraphics[width=10cm, angle=0]{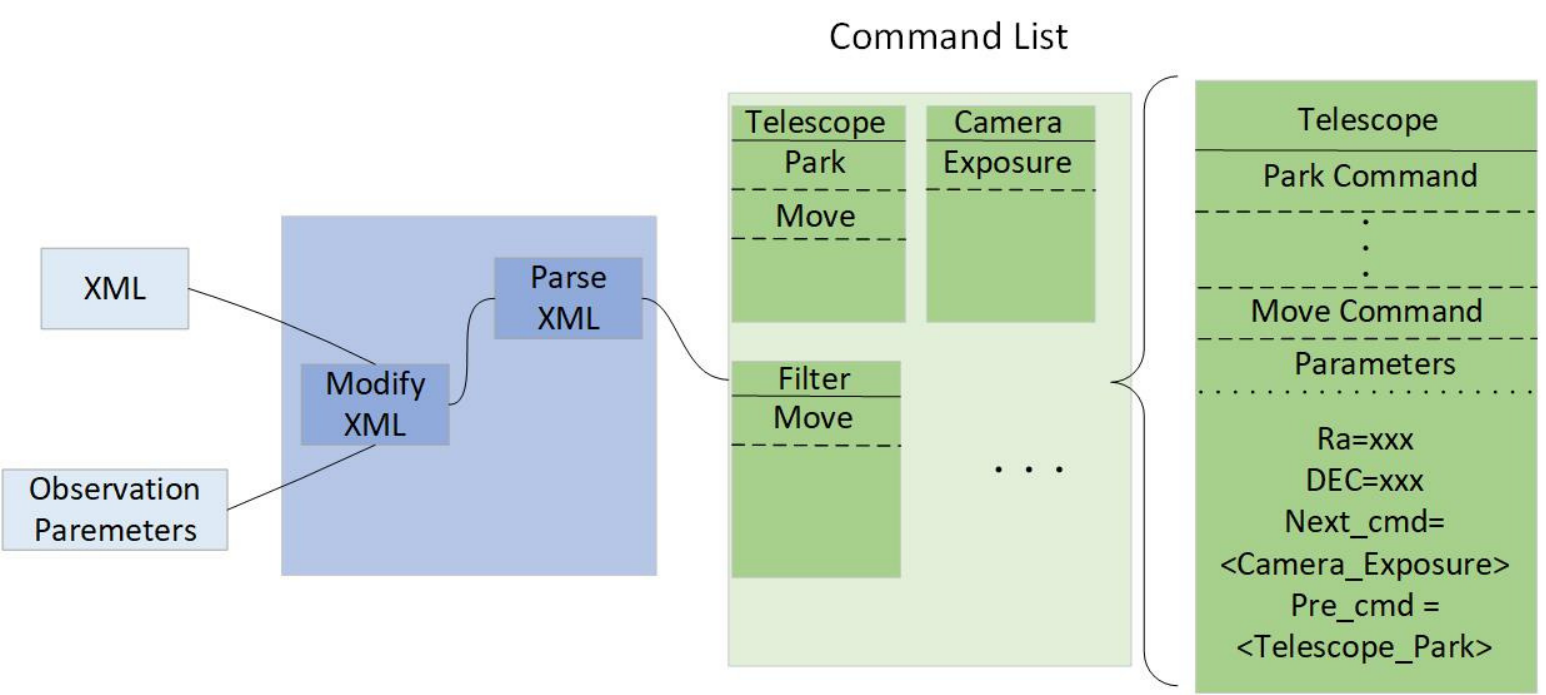}
\caption{Parsing process of the observation script}
\label{Fig2}
\end{figure}

The observation script only describes the control logic of the observed devices, which cannot be controlled directly. In the script parsing process (Figure 2), the scripts are first modified according to the observed target, and then parsed into a command flow queue with the creation time as the unique identifier. Each queue represents the command flow required by a device for sequential execution. For example, in the “Move” command of TCS, “pre\_cmd=Telescope\_park” indicates that the “Move” command is executed following the “Park” command, and “next\_cmd=Camera\_Exposure” indicates that the “Exposure” command of the camera is executed following the “Move” command.

\subsection{Control command executor}
The command executor is the core module of autonomous observation, and controls the execution sequence of the devices based on the command flow generated by the script parser. The command executor consists of a command flow queue, command task queue, and status monitoring module (Figure 3). The command task queue holds the tasks that are being executed currently or about to be executed by each device, while the status monitoring module receives the command execution status.

During command execution, a command is first selected in order from the command flow queue of the device list. If no dependent command is present or all commands in “pre\_cmd” have been executed, a command task is created in the task queue, and the commands are sent to a virtual machine. Further, the current task is suspended, awaiting resumption by the command execution status.

\begin{figure}[!htbp]
	\centering
	\includegraphics[width=12cm, angle=0]{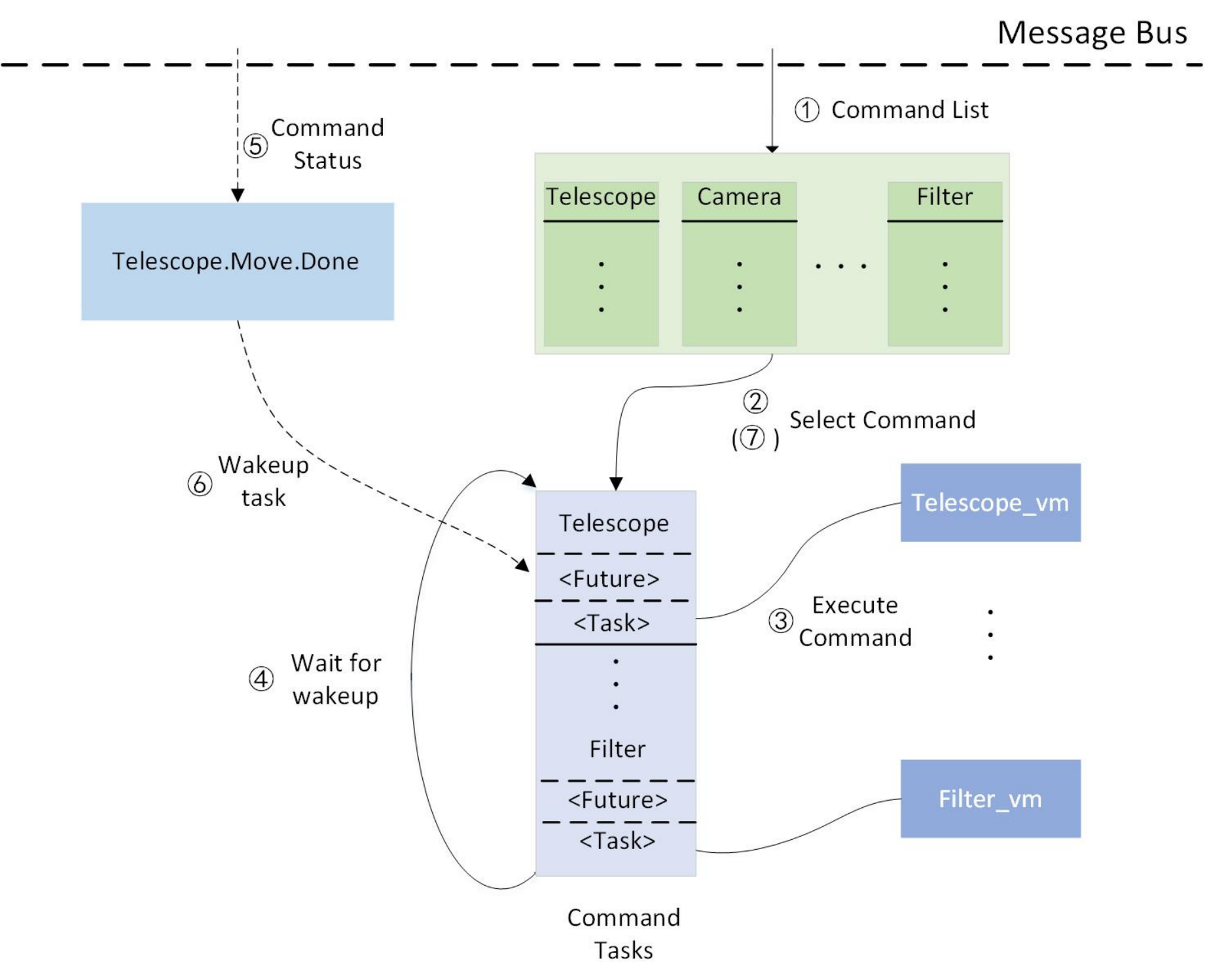}
	\caption{Schematic explaining the functions of the command executor}
	\label{Fig3}
\end{figure}

The command execution status is generated by the device virtual machine in real time and indicates the latest status of command execution. Figure 4 shows the binary command execution state code, which is used to represent all states in a command execution.

\begin{figure}[!htbp]
	\centering
	\includegraphics[width=8cm, angle=0]{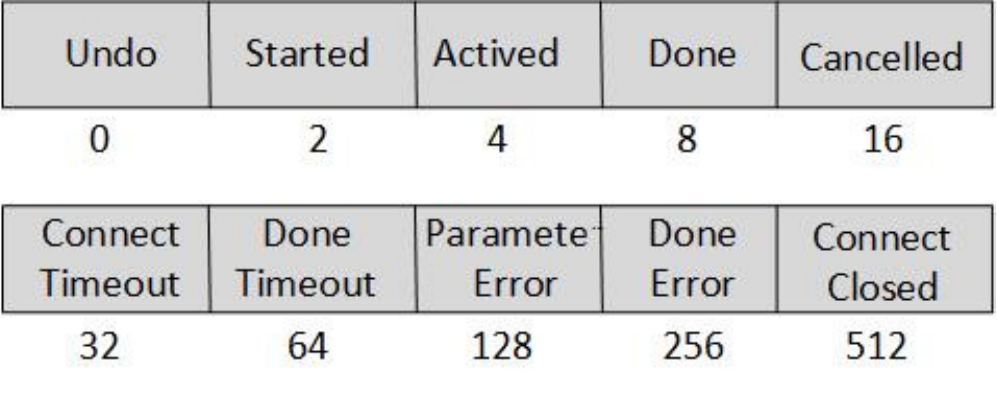}
	\caption{Binary command execution state code}
	\label{Fig4}
\end{figure}

Figure 5 shows the transition diagram of finite state automation for command execution, which can be used to monitor the command execution process. The initial state of each command is “Undo,” which changes to “Started” after the command is sent. “Actived” and “Done” indicate the start and completion of command execution by the device, respectively. The exception handling process of command execution is introduced later in Section 2.5.

\begin{figure}[!htbp]
	\centering
	\includegraphics[width=12cm, angle=0]{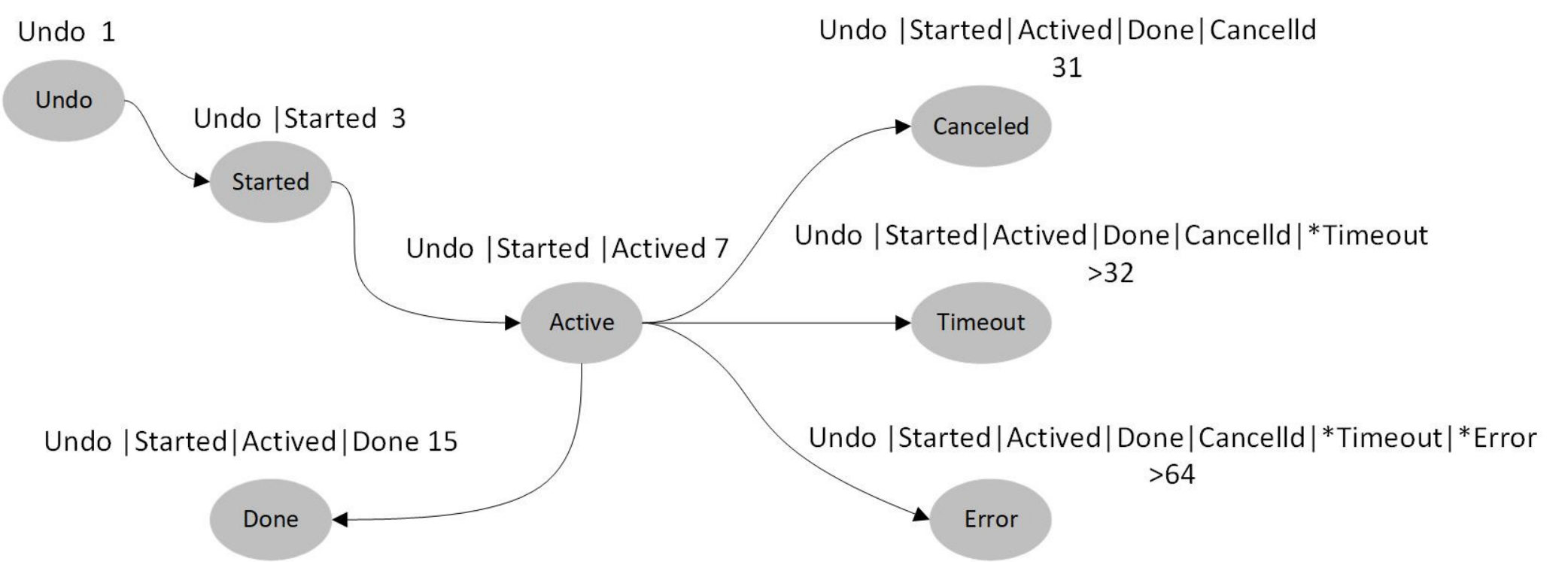}
	\caption{State transition diagram of finite state automation}
	\label{Fig5}
\end{figure}

Upon receiving the command completion status, the status module resumes the waiting task, changes the command state as “Done,” and executes the next executable command; this process is continued until all commands in the command queue are executed.

\subsection{Device virtual machine}
The device virtual machine is an independent unit that is directly connected with the devices. It receives commands and global status, converts the device control command format, distributes commands, generates command execution statuses, analyzes device exceptions, generates abnormal events, and alarms the UOCS. It is the core component of device control.

Although the execution processes, methods, and communication protocols of the command differ with each device, all undergo the four processes of translation, execution, monitoring, and logout. The “Device” base class implements the basic control logic of the commands and provides various rewritable interfaces, enabling the quick development of device virtual machines. For the connection mode of a device, it can be dynamically modified through the configuration file.

\begin{figure}[!htbp]
\centering
\includegraphics[width=\textwidth, angle=0]{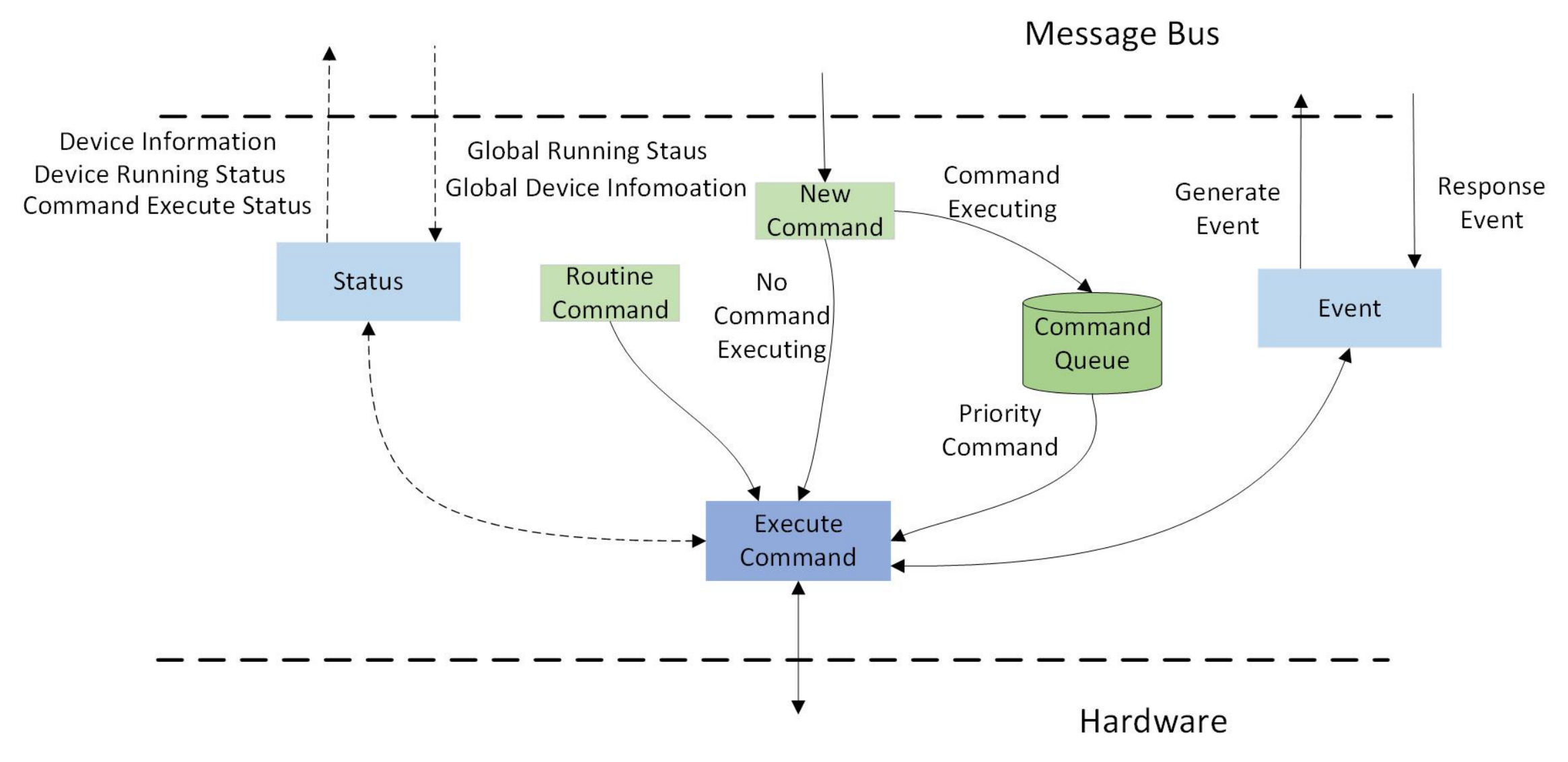}
\caption{Operation process of the device virtual machine}
\label{Fig6}
\end{figure}

Figure 6 shows the internal operational process of the device virtual machine. The virtual machine only receives one command in the automatic observation mode, but receives multiple commands simultaneously in the manual observation mode. The execution priority of commands could be low, medium, or high. For example, if a new command has lower priority than the current command, the new command is moved to the command queue. After the execution of the current command, the command with the highest priority is obtained from the command queue. On the contrary, if the new command has higher priority than the current command, the execution of the current command is stopped, and the new command is executed.

During the operation of the virtual machine, the real-time status of the device is regularly queried by a routine command, and then the running status and device status are sent to the status collector module. The running status is a real-time status of the virtual machine generated according to the command execution. The device status represents the detailed information of the device. Each command execution is a separate coroutine task. Figure 7 schematically describes the process of command execution in detail.

\begin{figure}[!htbp]
	\centering
	\includegraphics[width=\textwidth, angle=0]{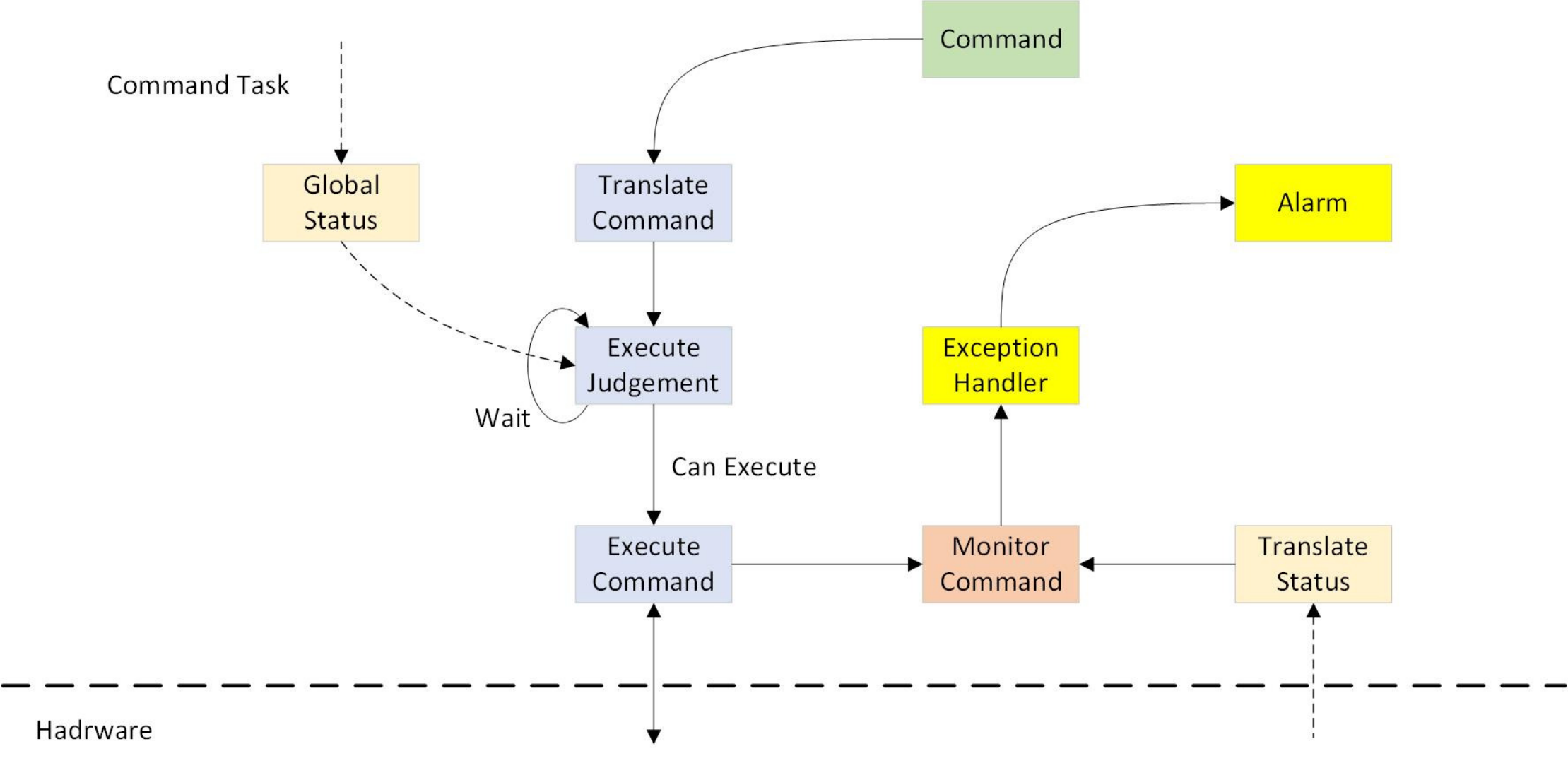}
	\caption{Command execution process of the device virtual machine}
	\label{Fig7}
\end{figure}

Although the command executor can automatically control the execution sequence of the devices according to the observation script, if the control logic is confused in the manual observation mode, observation errors could arise, affecting the observation efficiency. For example, in the camera exposure command, if the TCS moves during camera exposure, the target image would be distorted.

The global status is a set of running statuses of each module in the UOCS. The virtual machine determines if the command can be executed according to the global status, prior to beginning the execution. For example, the camera exposure command is executed only if the statuses of the TCS and the filter are respectively “tracking” and “ready”. In the command execution process, the command execution state is monitored based on the real-time status information of the device. In case exceptions occur, an exception event is generated (discussed in the next subsection).

\subsection{Exception handling mechanism}
Depending on the producer, exceptions could be either device exceptions or internal exceptions. Device exceptions are produced by the device module during command execution and are of three types: command delete exception, command execution timeout exception, and command execution exception. Internal exceptions include memory leakage and other exceptions that occur during the module operation.
Exceptions are first attempted to be solved within the module; if this is unsuccessful, an exception event is generated and sent to the message bus, which is captured by the user interface and alerted to the observer. The observer will either ignore, re-execute, abandon, or suspend the observation depending on the content and level of the exception.

\subsection{Status collector}
The status collection module collects, summarizes, collates, and distributes the status information of all modules in the UOCS.

\begin{figure}[!htbp]
	\centering
	\includegraphics[width=12cm, angle=0]{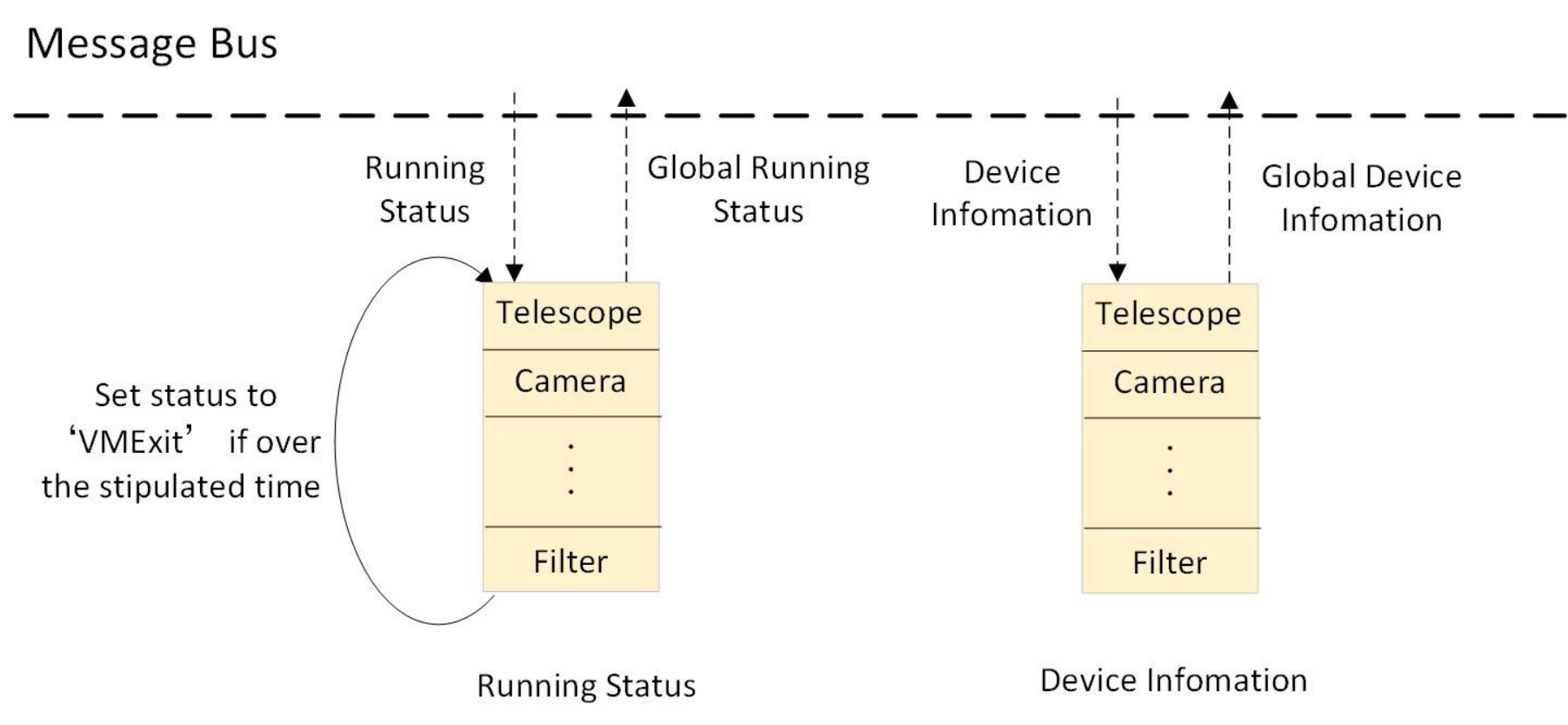}
	\caption{Process flow of status collection}
	\label{Fig8}
\end{figure}

Although the device running statuses in the UOCS differ depending on their functions, they have the same state characteristics. For example, all modules consist of an offline (VMExit) and waiting status (ready). The running statuses of all modules are set to “VMExit” when the status collection module starts and are updated and sent to the message bus in real time upon receiving the latest running status of each module. When a module exits, the status collection module sets the module running status to “VMExit” and generates a “module exit event” to notify all modules in the UOCS.

\section{Testing of UOCS performance}
\label{sect:Testing}
Following the completion of the UOCS design, we deployed the UOCS in the LAMOST working environment for simulation observation. The UOCS server runs on four cores (including eight logical cores) and has a memory of 32 GB. The device client simulates 50 devices through five computers, each running on an Intel® Core™ i3-2100 Processor (clockspeed: 3.10 GHz) with 4 GB of memory. All computers were connected through a 10 Gb switch and used the CentOS7 operating system. To verify if the performance of the UOCS communication system can meet the requirements of astronomical observation and control, a series of communication performance tests were conducted. 

\subsection{Performance of communication bus}
\subsubsection{Performance of message bus}
The device client randomly generates data (8 B, 16 B...1 kB) composed of alphanumeric characters and sends them to the message bus. The next data are sent after receiving the current data returned by the message bus. To ensure measurement accuracy, each data unit is sent 1 000 000 times repeatedly; the test results are presented in Figure 9.

\begin{figure}[!htbp]
	\centering
	\includegraphics[width=10cm, angle=0]{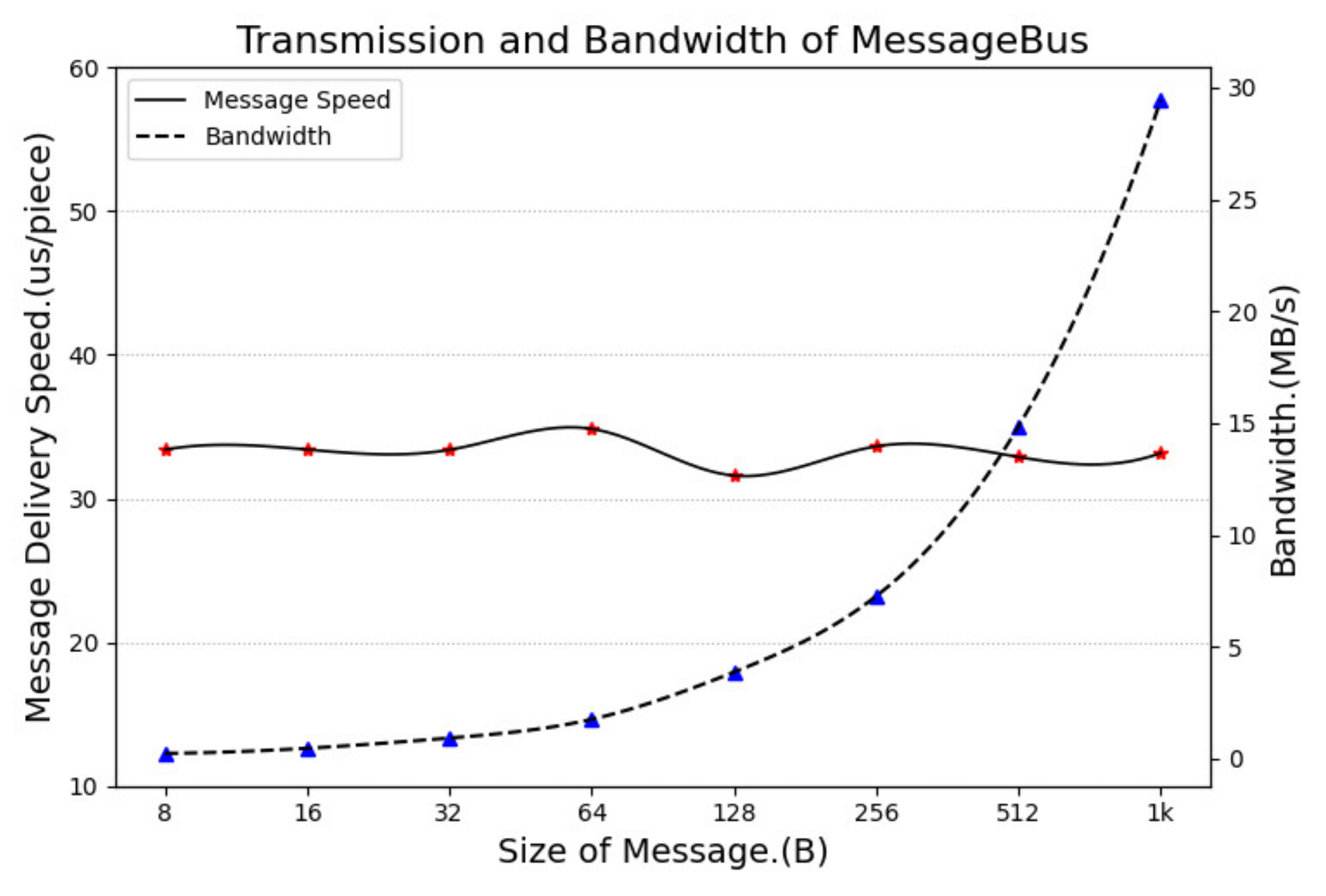}
	\caption{Transmission performance of the message bus for messages of different sizes}
	\label{Fig9}
\end{figure}

The message bus takes 33 us on average to deliver the received message. Although the message delivery time prolongs or shortens based on the data size, the difference is only ±5 us. The network bandwidth increases the with increasing size of the message packets, reaching a maximum of 28 MB/s.

\subsubsection{Performance of data bus}
The device client randomly generates (1 MB, 8 MB...128 MB) image data and sends them to the data bus. The next image is sent after receiving the data returned by the data bus.To ensure measurement accuracy, each image is repeatedly sent 10 000 times; Figure 10 presents the test results.

\begin{figure}[!htbp]
	\centering
	\includegraphics[width=10cm, angle=0]{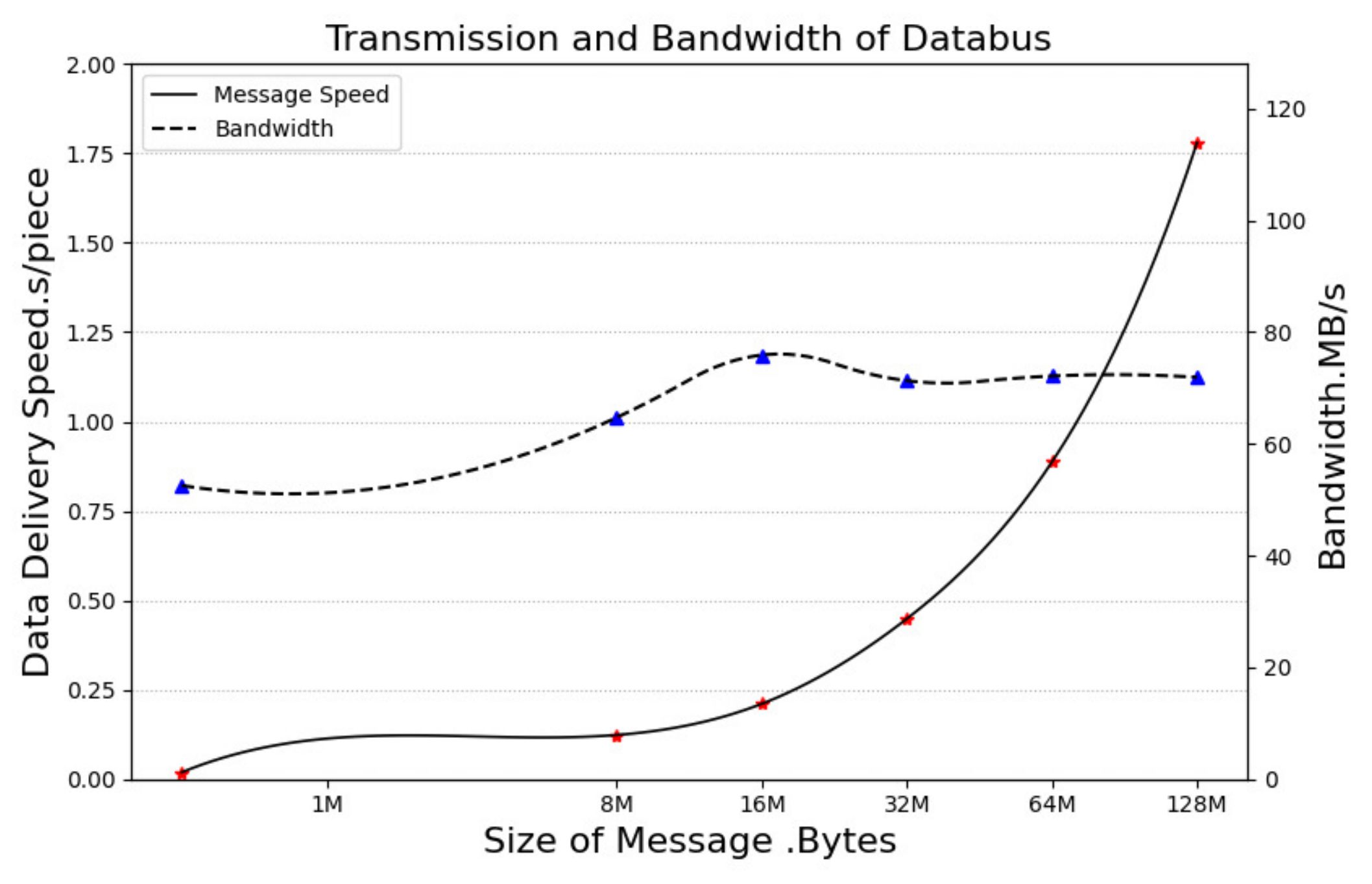}
	\caption{Transmission performance of the data bus for images of different sizes}
	\label{Fig10}
\end{figure}

As seen from the figure, the data delivery time increases with increasing size of the data packets. The longest time of 1.75 s was required to deliver the data of size 128 MB. The network bandwidth reaches a maximum of 75 MB/s for the data packet of size 16 MB, and subsequently remains around 71 MB/s. This phenomenon occurs because the internal cache of the data bus is full, resulting in a delay in forwarding the data. 

\subsubsection{Influence of data bus on message bus}
The data bus and message bus are two independent transmission modules. Nevertheless, when sharing the same network hardware, they influence each other because of hardware resource contention. Figure 11 shows the impact of such a phenomenon on message delivery, when the data bus and message bus are deployed on the same computer.

\begin{figure}[!htbp]
	\centering
	\includegraphics[width=10cm, angle=0]{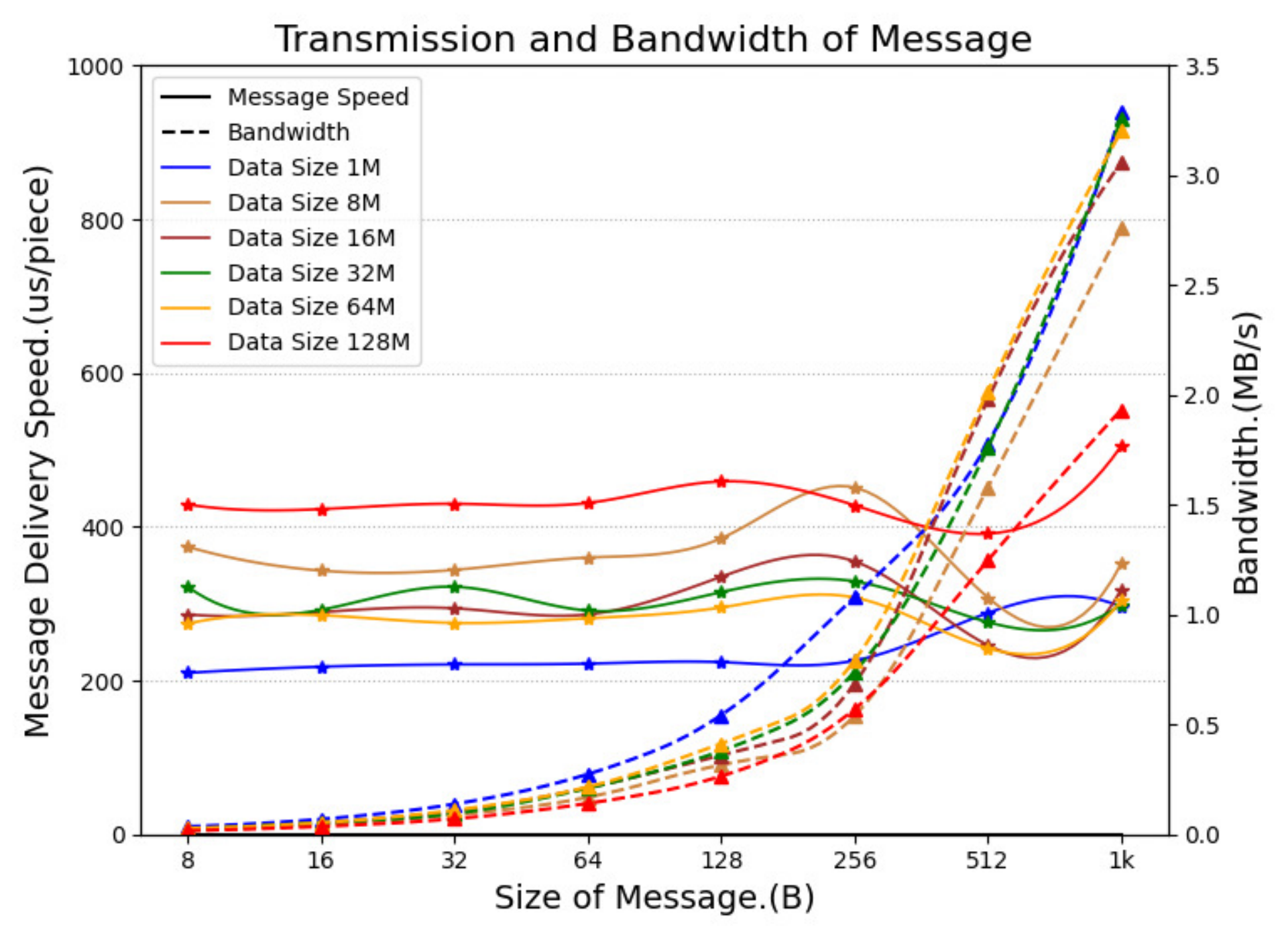}
	\caption{Performance of message bus under different sizes of data transmitted by the data bus}
	\label{Fig11}
\end{figure}

For the image transmission on the data bus, compared with the independent transmission on the message bus, the message delivery time increased 10-fold. The delivery time differed for images of different sizes. For an image of size 1 MB, the message delivery time was approximately 200 us. The maximum delivery time for a single message was approximately 500 us, for transmitting an image of size 128 MB. 

\subsection{Performance of status collector}

The status collector receives and delivers the latest status information. Note that the integrity and real-time collection of status information influence the stability of the entire system. The device client generates two types of information about the detailed status and the running status of the device in a certain time interval (1 ms, 2 ms... 10 ms), and it delivers the information to the status collector. The size of the running status information of the device is fixed at 8 B, whereas that of the detailed status of the device differs, e.g., 8 B, 16 B...1 KB. Figure 12 shows the performance of the status collector at different status delivery intervals.

\begin{figure}[!htbp]
	\centering
	\includegraphics[width=10cm, angle=0]{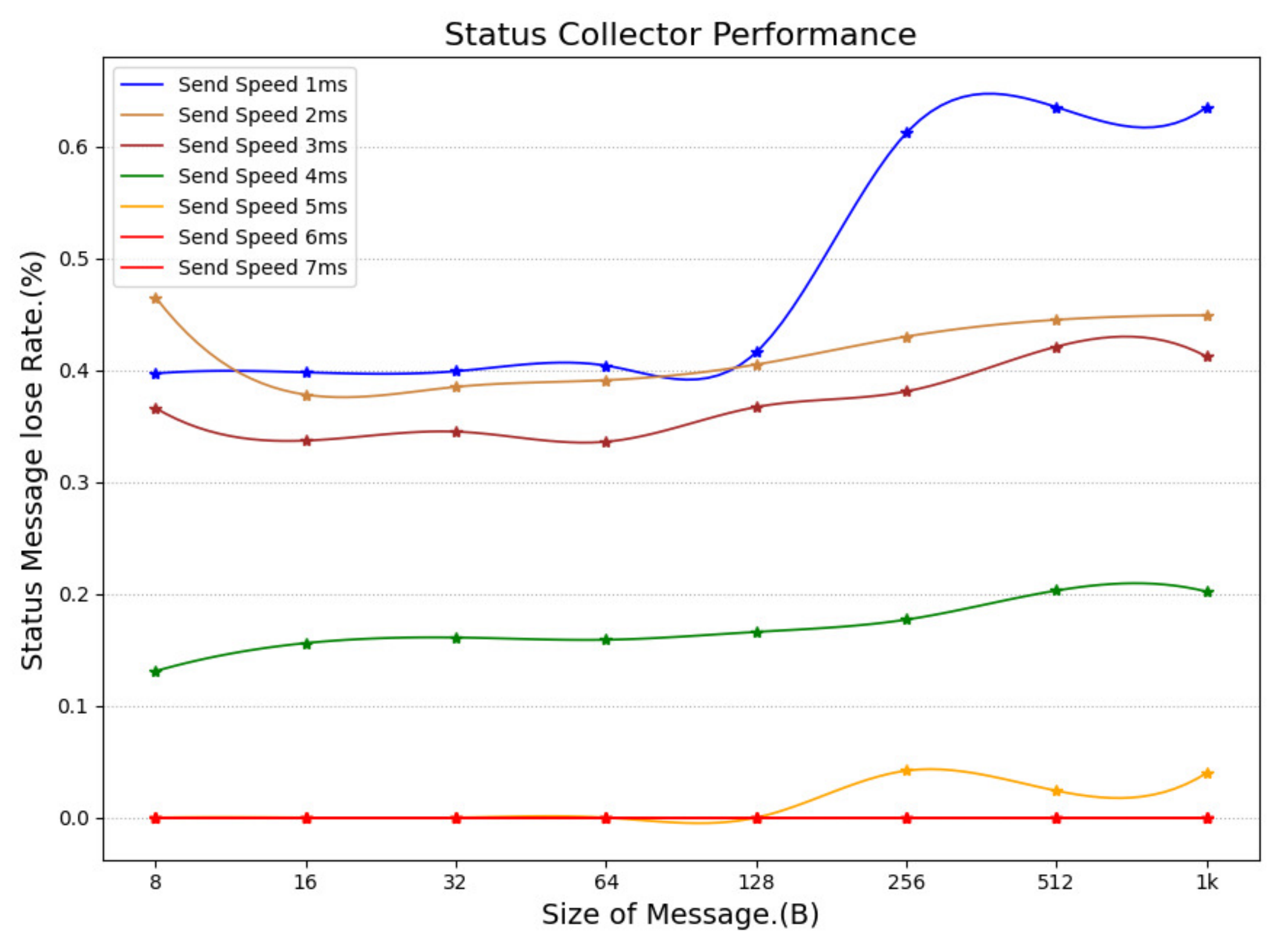}
	\caption{Performance of the status collector at different status delivery intervals}
	\label{Fig12}
\end{figure}

As shown in Figure 12, the packet loss is closely related to the status delivery interval and packet size. If the delivery interval is 6 ms or longer, no packet loss occurs. However, if the interval is shorter than 6 ms, the packet loss rate increases with increasing packet size. The maximum packet loss rate of 63\% occurs when the interval is 1 ms and packet size is 1024 B. 

The status collector collects the status information of only the subsystems and device control systems. For example, the active optical control system of LAMOST sends only the current key information to the status collector, and does not contain the real-time information of thousands of controllers. Furthermore, the status transmission interval of the subsystems and the device control system is substantially longer than the millisecond level, which is basically secondary. Therefore, the status collector of the UCOS fully meets the demands of the current telescope for status collection. 

\section{Application of UCOS to LAMOST}
\label{sect:application}
During the first phase of survey of LAMOST (2011–2017), more than nine million low-resolution spectra were obtained\footnote{http://dr5.lamost.org/}. At the beginning of the second phase, the medium-resolution survey was included, and more than 2 million low-resolution spectra and medium-resolution spectra have been obtained since 2017\footnote{http://www.lamost.org/dr8/}\footnote{http://dr7.lamost.org/}. According to the observation plan, the plate will be exposed two or three times, with an exposure time of 20–30 min for low-resolution spectra and 10–20 min for medium-resolution spectra.

\subsection{Observation flow of LAMOST}

The LAMOST comprises several subsystems (Figure 13), including the data handling system (DHS), survey strategy system (SSS), telescope control system (TCS), instrument control system (ICS), auxiliary monitoring system (AMS), and observation control system (OCS). The subsystems comprise one or several systems or devices, which are independent of each other and deployed in different control computers. For example, TCS consists of the mounting of telescope (Teld), focal plane (FP), dome, guiding system (Guide), and active optics System (AO), among others. The ICS system includes the camera cluster control system (Camera) and optical fiber control system (Fiber); the DHS processes the images obtained from cameras(\cite{luo2010}). Plates are generated by SSS according to the sky survey and provided to astronomers for selection. Each plate comprises several files, including those on the central star, guide star, 4000 targets, active optical parameters, and other observation information.

\begin{figure}[!htbp]
	\centering
	\includegraphics[width=\textwidth, angle=0]{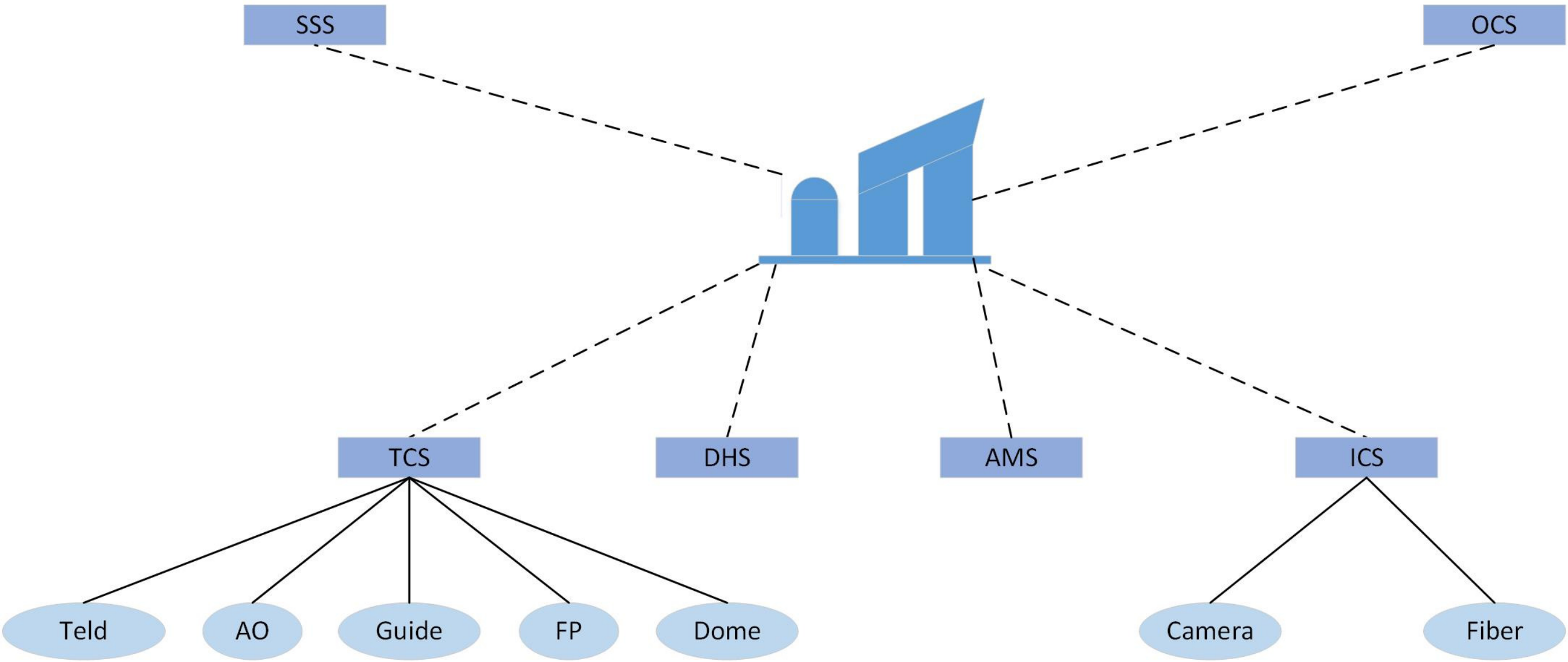}
	\caption{LAMOST structure: subsystems and devices(DHS:Data Handling System;AMS:Auxiliary Monitoring System;AO:Active Optics System;FP:Focal Plane;SSS: Survey Strategy System;TELD:Telescope)}
	\label{Fig13}
\end{figure}

During observation, the LAMOST is operated in collaboration with the guide operator, Teld and FP operators, AO operator, and ICS and DHS operators. Figure 14 shows the control flow of LAMOST observation. Each block represents an operation and contains information about the name, execution time, and steps required to complete the operation. The horizontal operation is sequential, whereas the vertical operation is a parallel operation.

\begin{figure}[!htbp]
	\centering
	\includegraphics[width=\textwidth, angle=0]{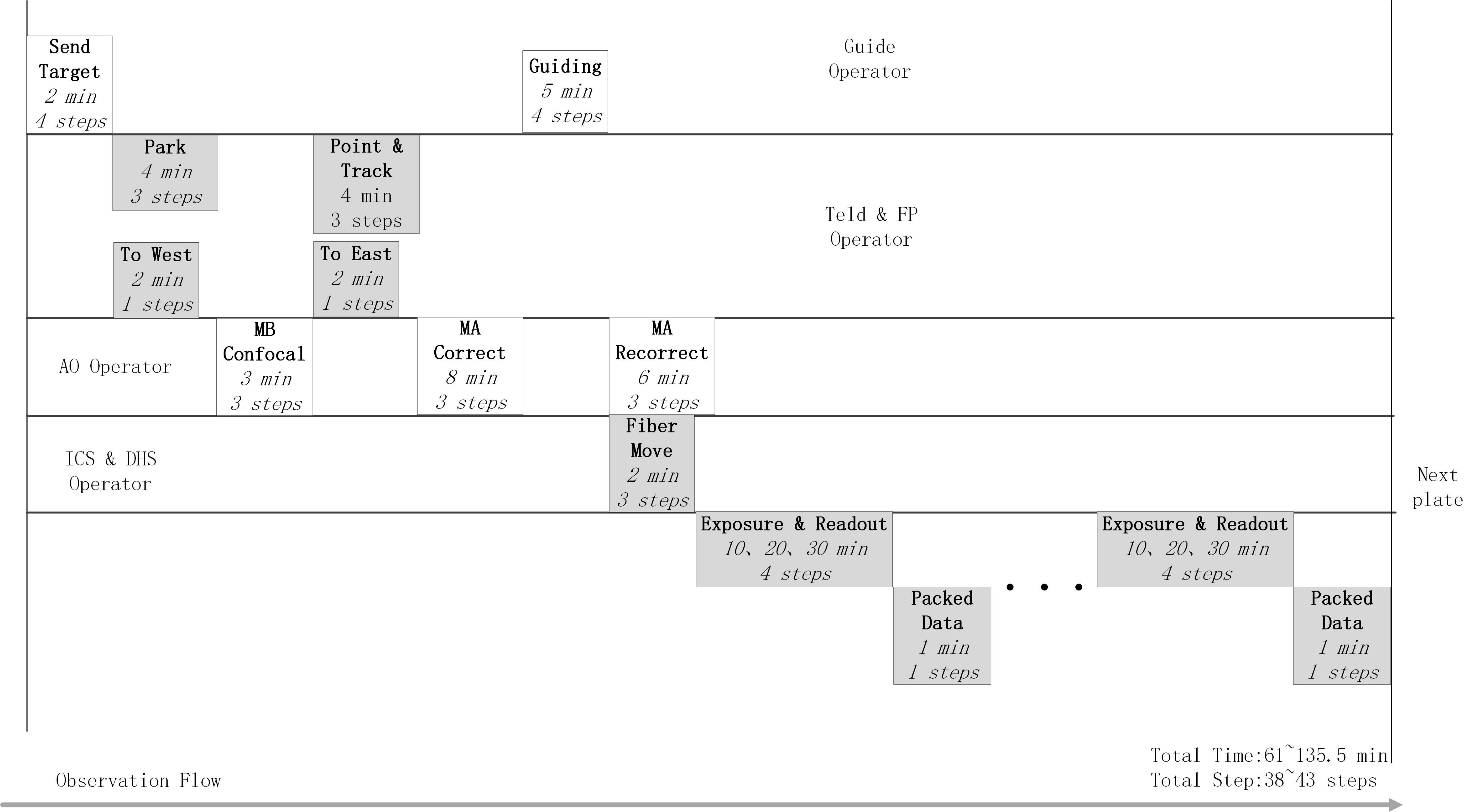}
	\caption{OCS Control Flow}
	\label{Fig14}
\end{figure}

Before beginning the observation, the astronomer selects the required observation plate from the plate list based on the meteorological conditions and observation plan, and notifies all the operators to commence observation. Successively, the guide operator sends the plate information to all subsystems, following which other operators perform corresponding operations, such as pointing of Teld toward a specific bright star, moving away of FP, MB confocal, Teld pointing and tracking, FP moving back and tracking, MA correction, guiding, MA re-correction, fiber moving and tracking, camera exposure, and image packaging. The time required for a plate differs with the survey type. For example, in a medium-resolution survey, for two exposures (wherein each exposure lasts 10 min), it is 61 min; for three exposures (wherein each exposure lasts 20 min), it is 105.5 min. In a low-resolution survey, for two exposures (wherein each exposure lasts 20 min), it is 81 min; for three exposures (wherein each exposure lasts 30 min), it is 135.5 min. Furthermore, regarding the number of operation steps for a plate, the operator requires 38 steps for two exposures, and 43 steps for three exposures. During observation, the operators orally update each other of the operation progress, entailing 5\% communication errors, which must be rectified by repeated confirmation by the operators.

\subsection{UOCS application to LAMOST}
To ensure a stable and reliable observation of LAMOST, the UOCS is implemented in individual steps according to the principle of “simple first, complex second.” First, the automated control between modules is achieved, which surmounts the communication obstacles between the subsystems and devices. Subsequently, the interfaces of all subsystems and devices with UOCS are converted, ultimately realizing automated observation control of the LAMOST. 

\subsubsection{Selection of communication bus}
Currently, LAMOST comprises 10 devices or subsystems that must be connected with the UOCS. Among these, 32 cameras are controlled by the camera control system(~\cite{tian2018}); each camera generates 32 MB-images, resulting in a total image size of 1024 MB. Recall from Figure 11 that for the data of size 32 MB transmitted by the data bus, the delivery time of the message bus is 300 ms, which fully meets the requirement of the LAMOST in terms of the timeliness of observation and control. Therefore, in LAMOST, we deploy the message bus and the data bus together in the UOCS server to reduce the complexity of the system.

\subsubsection{Control Flow of UOCS}
The control flow of the UOCS observation script for LAMOST (Figure 15) is generated according to the control sequence between the subsystems and devices by comprehensively analyzing the operation steps of the operators. Unlike the centralized control mode for subsystems in OCS, UOCS is a distributed control mode for devices. In the figure, blocks of each color represent a command to be executed by the device. The script generator converts it into an XML file, while the script parser generates the control command stream. Each block is a device control command containing the device name, command name, and parameters, among others. Two sequential commands are connected by an arrow, the beginning of which represents the current command, whereas the pointing end represents the next command to be executed. A command list represents a serial command queue, which contains the commands that are executed sequentially. Different lists are parallel, which indicates that these commands can be executed simultaneously.

\begin{figure}[!htbp]
	\centering
	\includegraphics[width=8cm, angle=0]{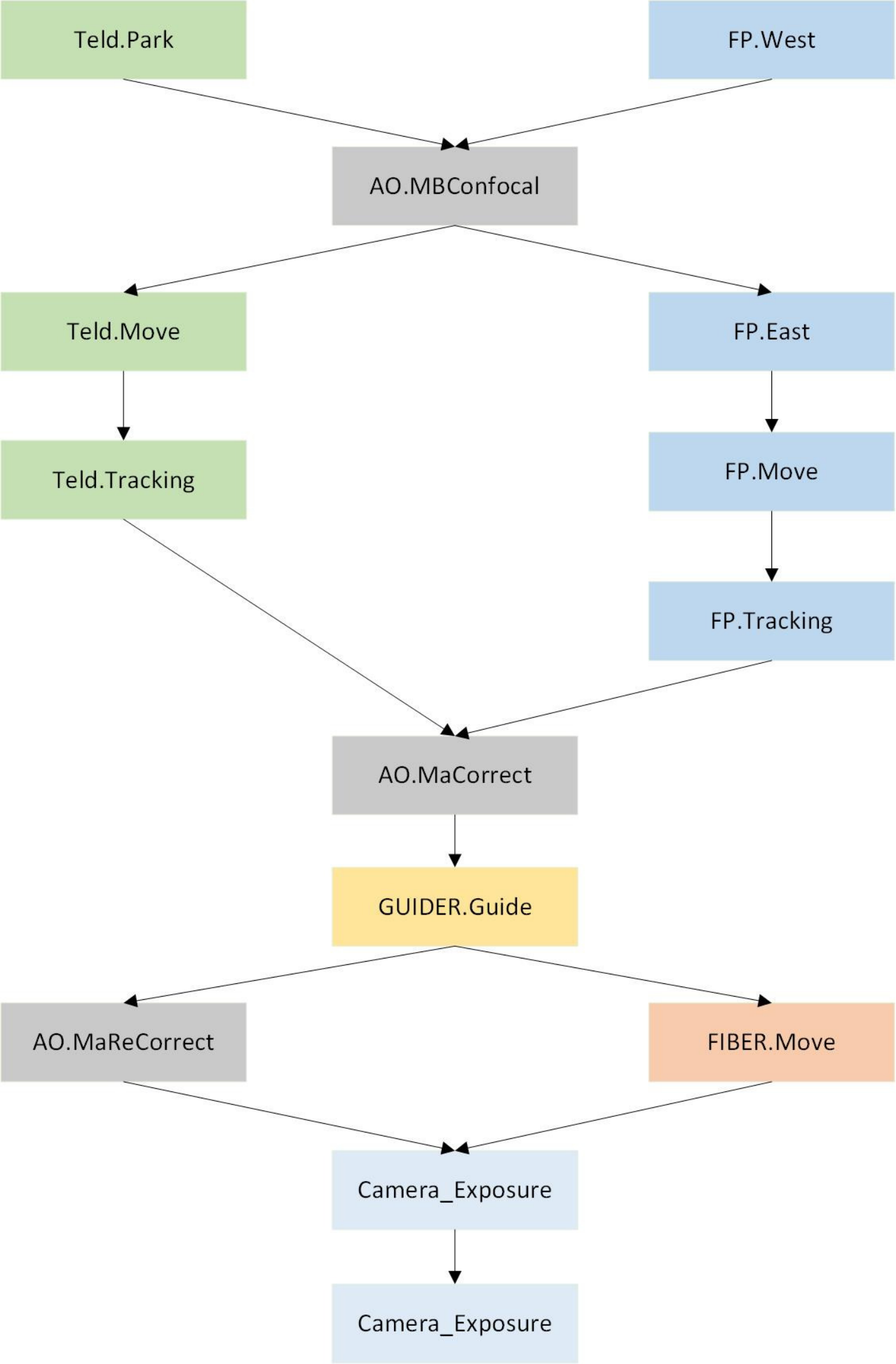}
	\caption{Flowchart of LAMOST observation script}
	\label{Fig15}
\end{figure}

Figure 16 shows the control logic of UOCS observation, which is implemented via device commands and event mechanisms. The dashed-line in the figure represents the event control flow between modules, while the solid line represents the control flow between the command executor and devices or subsystems.

\begin{figure}[!htbp]
	\centering
	\includegraphics[width=\textwidth, angle=0]{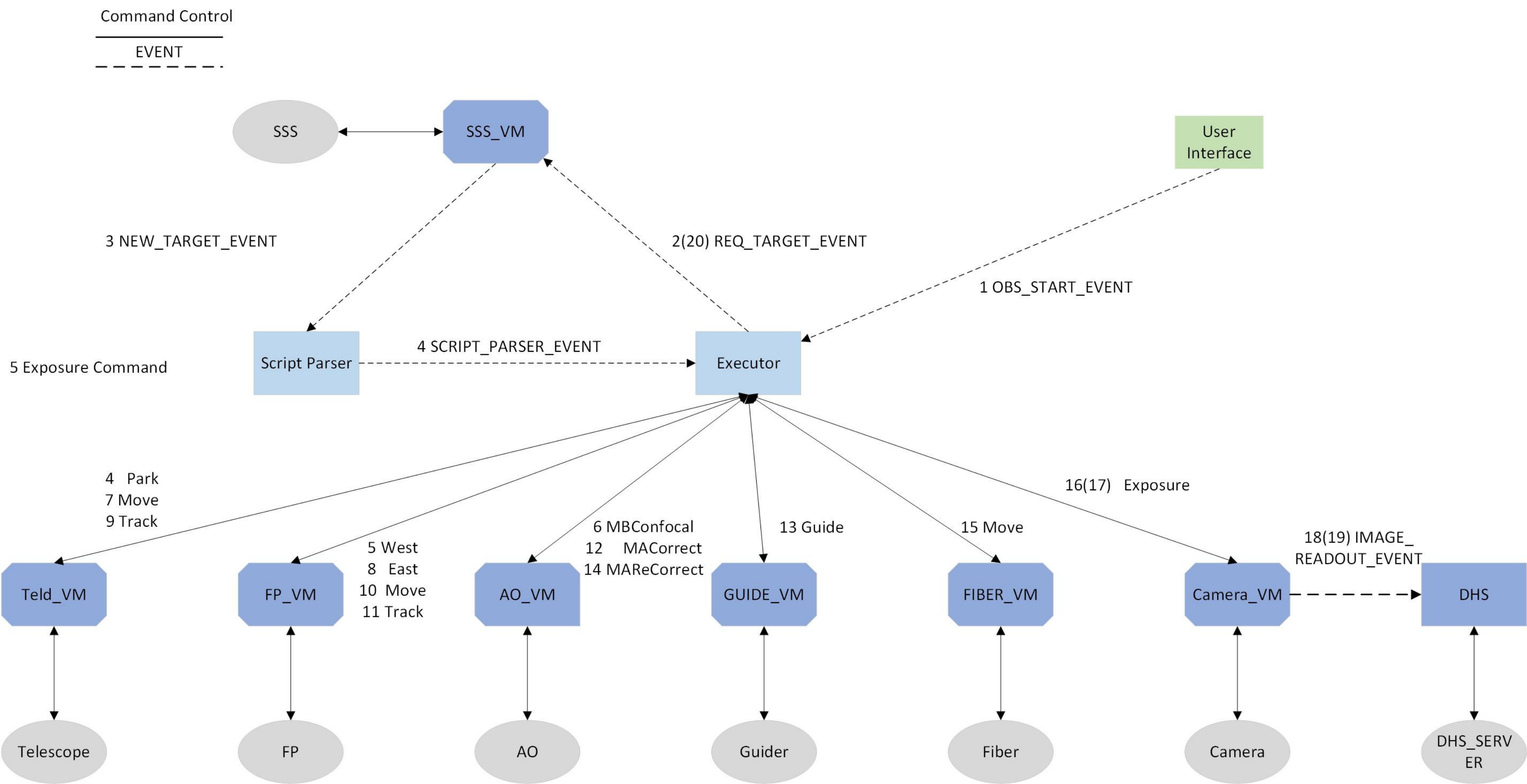}
	\caption{Control logic of LAMOST observation}
	\label{Fig16}
\end{figure}

The OCS and other subsystems are developed by different units. Despite the existence of a unified communication protocol, to realize automated control through UOCS, the control details of the subsystems and devices must be first understood, and then a perfect exception handling and control mechanism must be designed—this process requires a long time to study and prepare. The camera control system was maintained by the UOCS development team consisting of personnel familiar with the communication interface and execution logic. The fiber control system contains fewer commands, a simple logic, and less manual intervention. Therefore, automated control of these two subsystems of the UOCS was first realized; the automated control of other subsystems and devices will be gradually realized in the future.
 
Because the internal communication protocol within UOCS differs between the camera and fiber control, we used the device agent shown in Figure 17 to perform conversion for subsystems and devices.

\begin{figure}[!htbp]
	\centering
	\includegraphics[width=6cm, angle=0]{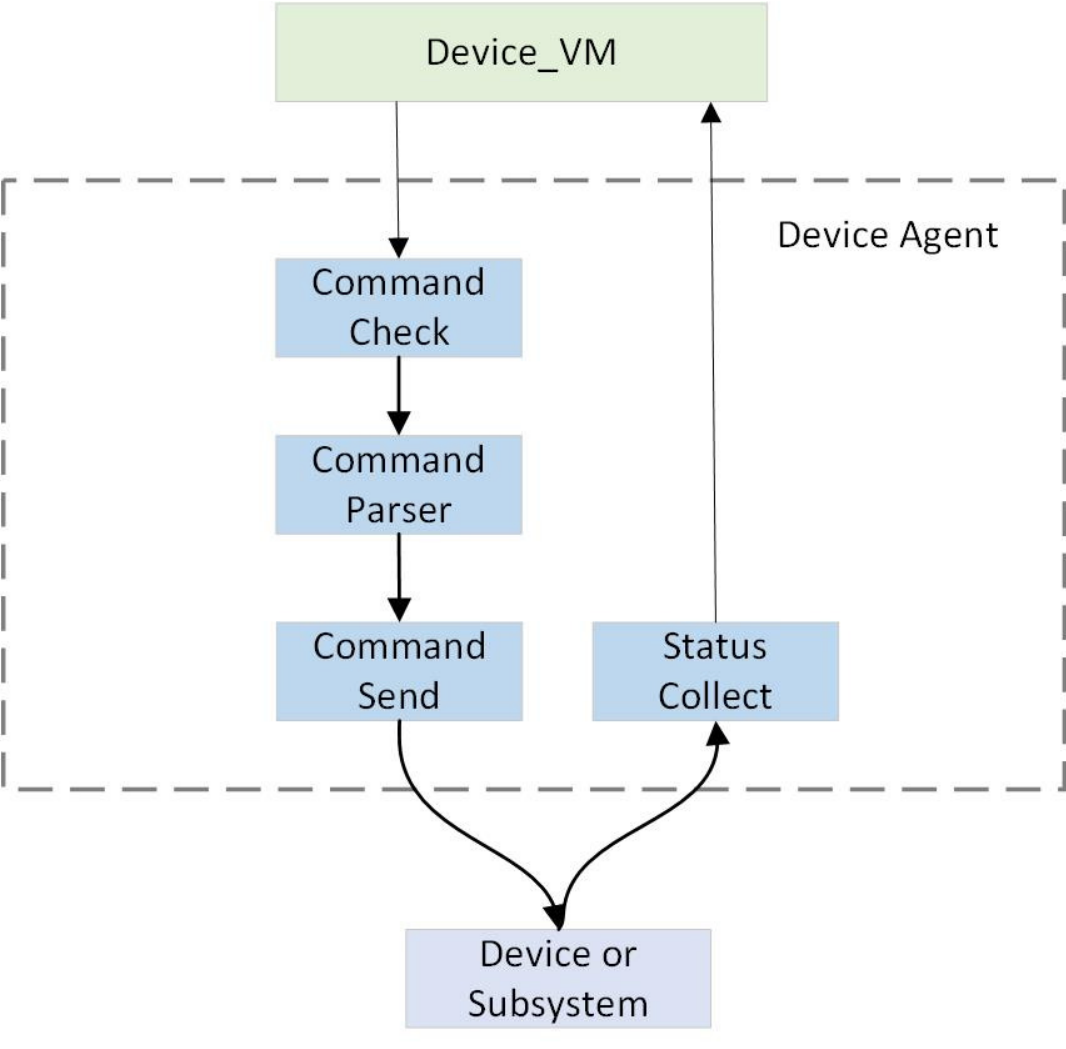}
	\caption{Device agent used to perform conversion for subsystems and devices}
	\label{Fig17}
\end{figure}

\subsubsection{Camera agent}
The camera agent involves four control commands: “Exposure,” “Abort,” “Pause,” and “Readout.” Each command involves multiple operation steps. For example, “Exposure” requires four steps: “Ready,” “Setup Parameters,” “Start Exposure,” and “Close.” In the UOCS observation script, the camera contains only one command for control exposure. If the four-step control method is adopted, the control script of UOCS not only becomes redundant, but also increases the complexity of the control logic of the command executor, making the whole system unstable. Therefore, the exception handling and camera control tasks are assigned to the camera agent in the UOCS. “Camera\_VM” is used to send a control command to the camera agent, which divides the command into several sub-control commands according to the camera protocol and sends them to all cameras. Additionally, the camera agent converts the cameras status into the internal device status of UOCS and feeds it back to the status collector through “Camera\_VM.”

\subsubsection{Fiber agent}
Fiber control involves three commands: “power-on,” “fiber movement,” and “power-off.” Upon receiving the “start observation” command, the script parser loads the information of 4000 fibers as parameters into the “Move” command, which is sent to “Fiber\_VM” by the command executor. Further, “Fiber\_Agent” sends the above-mentioned three commands to the fiber control system. 

\subsubsection{Other subsystem agents}
To ensure the implementation of the entire observation process in UOCS, the control commands for other subsystems and devices are received by deploying a virtual machine in the control computer. The virtual machine displays a dialog box with voice notification to prompt the operator to execute the corresponding operation. Upon completing the execution, the operator selects “Done” or “Error” according to the operation result and returns the command execution status to the executor.

\begin{figure}[!htbp]
	\centering
	\includegraphics[width=4cm, angle=0]{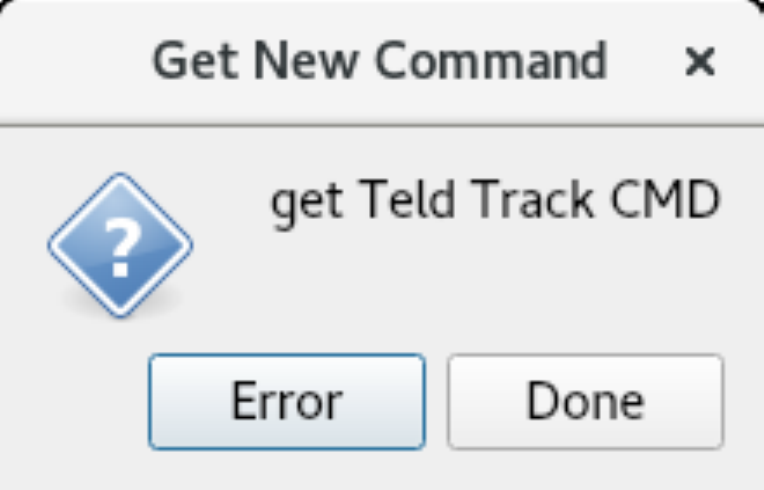}
	\caption{ “Teld” virtual machine dialog box}
	\label{Fig18}
\end{figure}

\subsection{Comparison of UOCS and OCS}

Figure 19 shows the control interfaces of the UOCS and OCS. In OCS (figure a), all information is displayed in the same interface; therefore, details on the operation information of all devices or subsystems are not available. For example, the information of only the overall progress of the cameras is displayed, while the operation details of each camera are hidden. On the other hand, the UOCS (figure b) provides a dedicated control and status display interface for each device.

\begin{figure}[!htbp]
	\centering
	\subfigure[Control Interface of OCS]{
	\begin{minipage}{6cm}
		\centering
		\includegraphics[width=10cm]{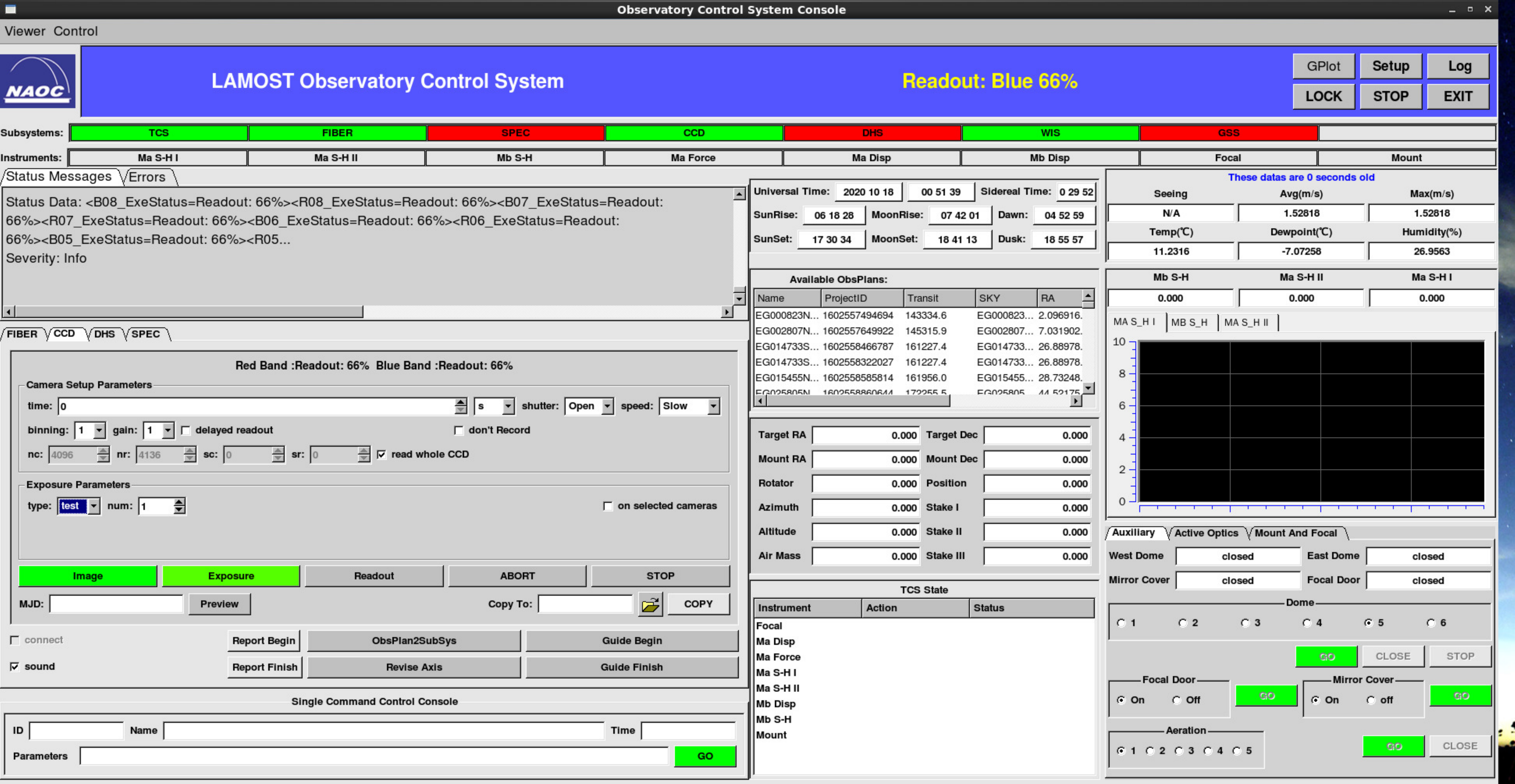}
	\end{minipage}
	}

	\subfigure[Control Interface of UOCS]{
	\begin{minipage}{6cm}
		\centering
		\includegraphics[width=10cm]{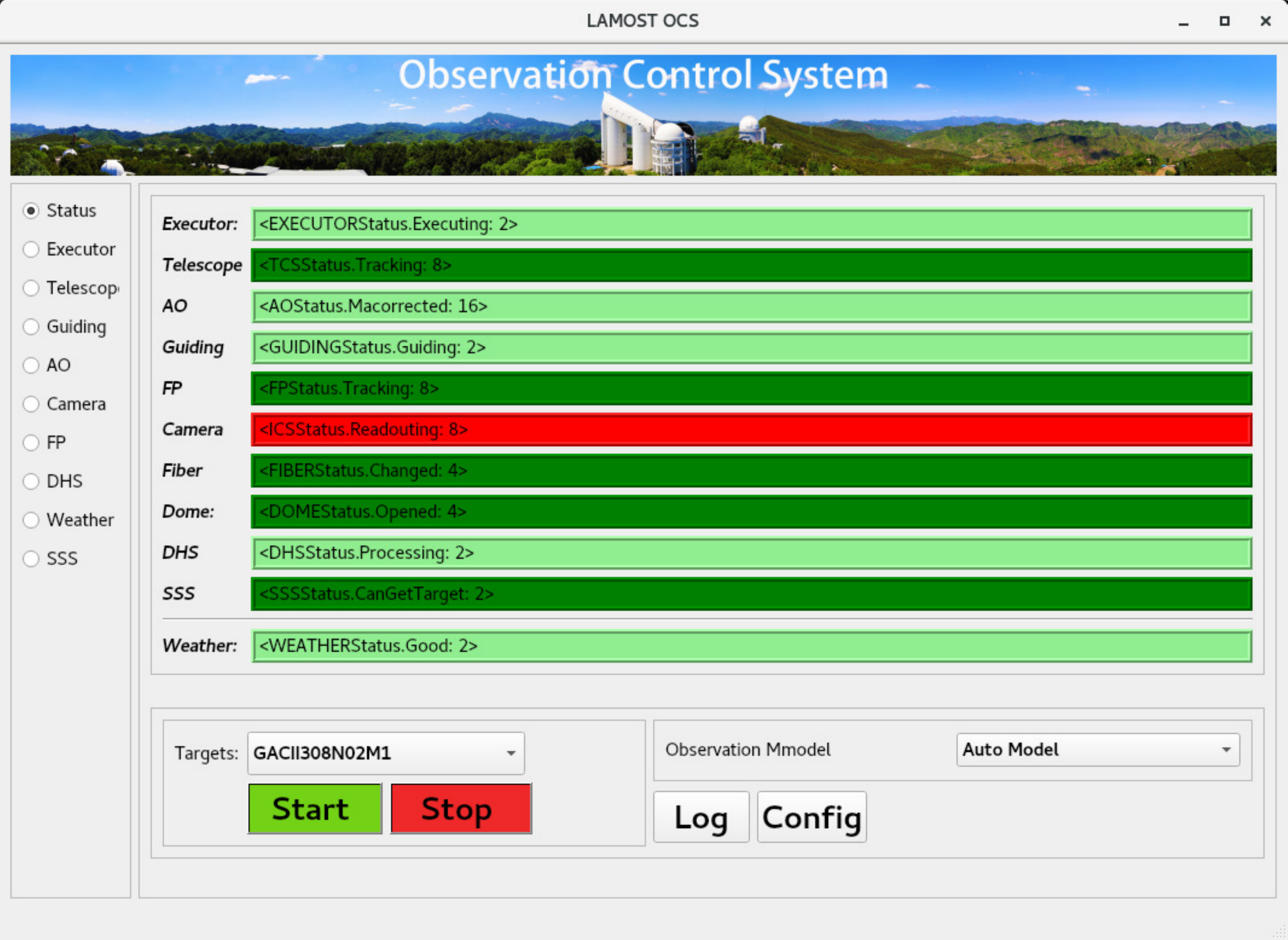}
	\end{minipage}
	}
	\caption{Comparison of the control interfaces of UOCS and OCS}
	\label{Fig19}
\end{figure}

Figure 20 shows the camera control interface of UOCS, which displays detailed information of 32 cameras during exposure. Compared with that of OCS, the layout of UOCS is more rational, and it more clearly displays the running status of each module. Therefore, each device can be controlled independently, greatly improving the convenience for operation control and debugging.

\begin{figure}[!htbp]
	\centering
	\includegraphics[width=8cm, angle=0]{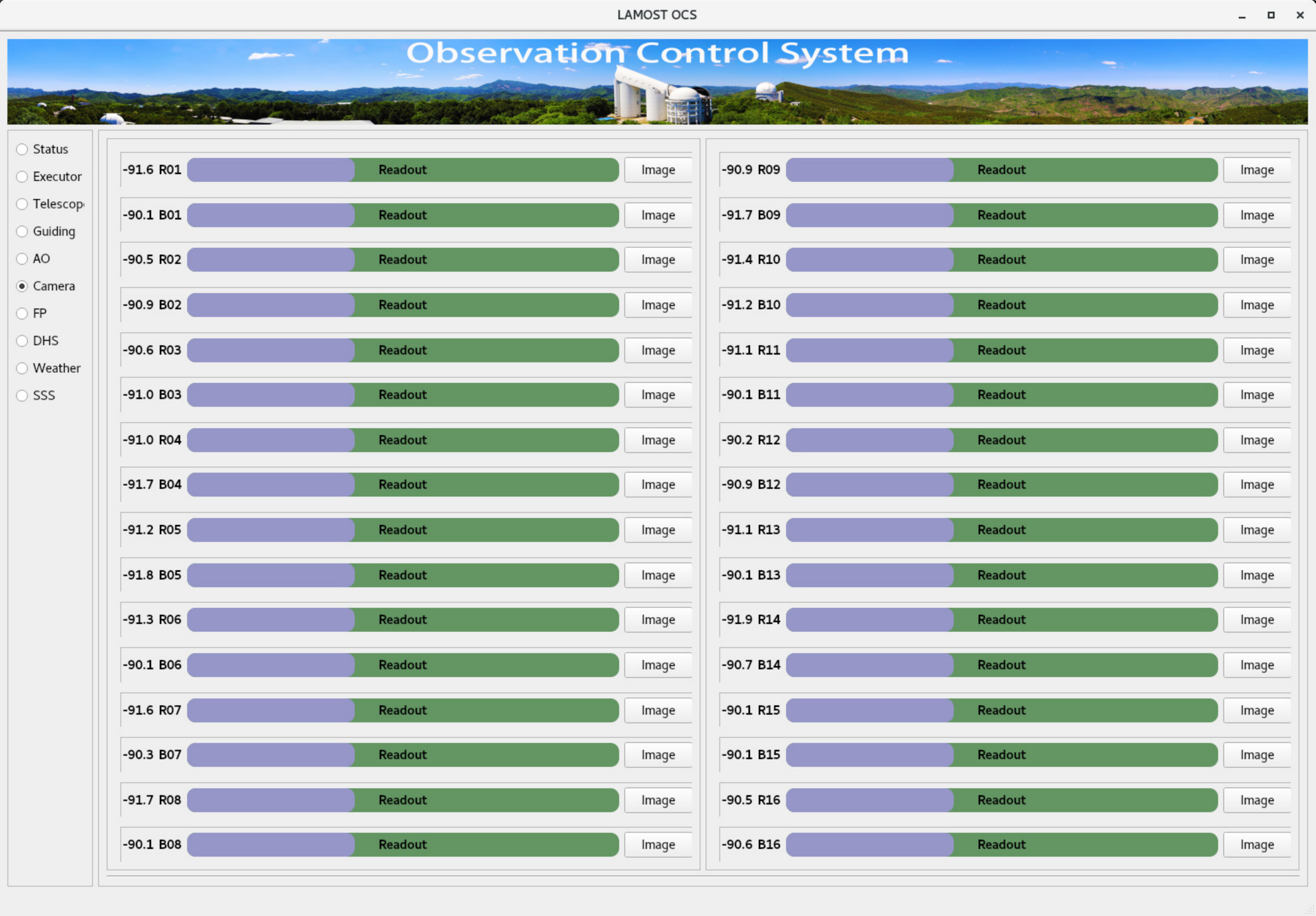}
	\caption{Camera control interface of UOCS}
	\label{Fig20}
\end{figure}

During observation, in the case of OCS, the astronomer selects a plate, informs the guide operator to notify other operators of the observation target, and starts the next plate when all the images have been packaged. Figure 21 shows the observation control flow of UOCS. The UOCS automatically sends it to all devices or subsystems after the astronomer selects a plate; the next plate is started after the exposure of the last image. Optimizing these two aspects can reduce 6.5 min for each plate, and decrease the number of operation steps by 15 for two exposures and 19 for three exposures.

\begin{figure}[!htbp]
	\centering
	\includegraphics[width=\textwidth, angle=0]{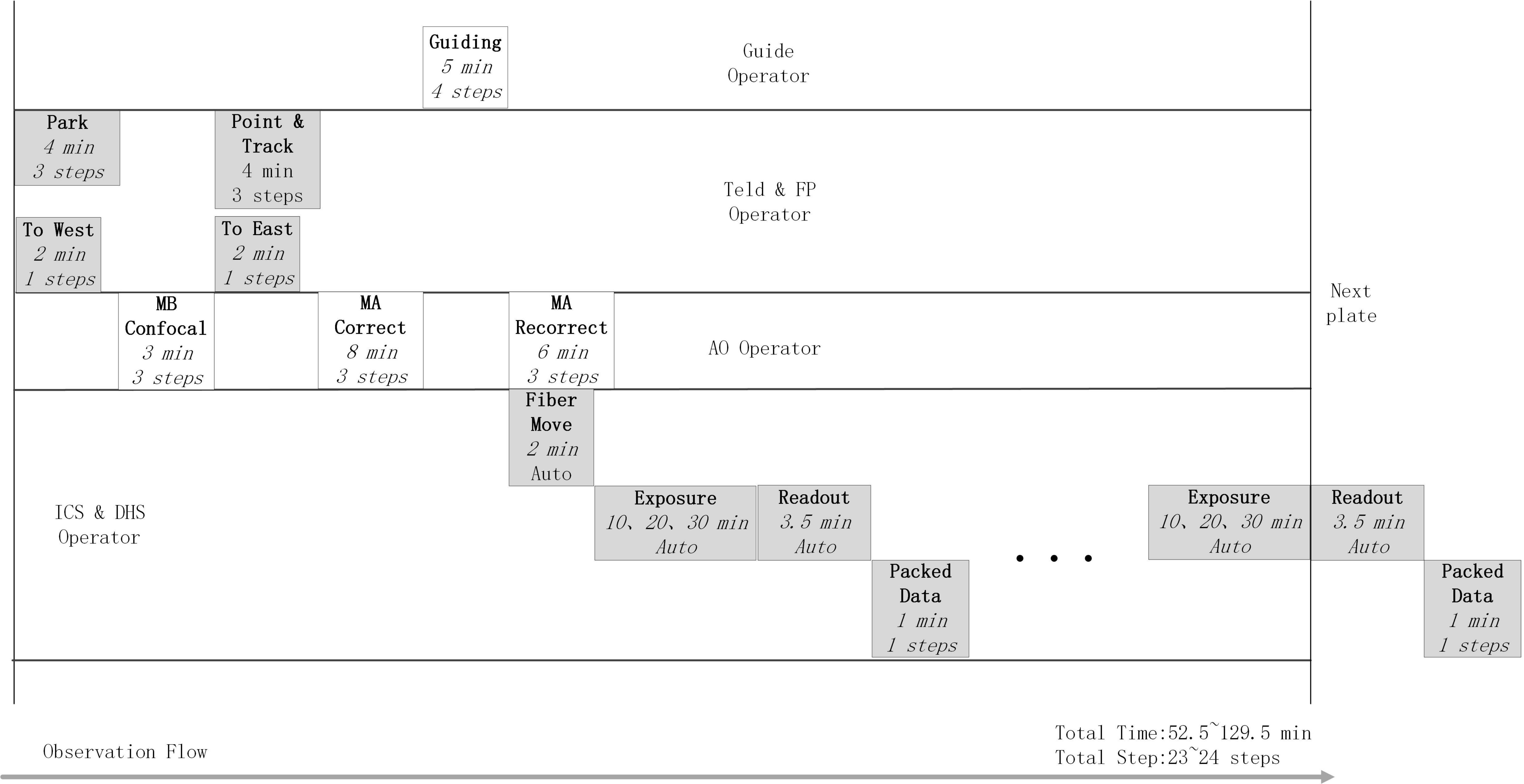}
	\caption{Observation control flow of UOCS for LAMOST}
	\label{Fig21}
\end{figure}

The UOCS performs significantly better than OCS in terms of the observation performance, operator complexity, and communication error. The efficiency of UOCS in medium-resolution surveys improved by a minimum of 6.8\% and maximum of 10.6\%, and that in low-resolution surveys improved by a minimum of 4.8\% and maximum of 6.2\%. Further, the operational complexity of UOCS reduced by 39.4\% for two exposures and 44.2\% for three exposures. In UOCS, operators are no longer required to orally update each other of the operation progress, because the operations required for the subsystems and devices are automatically broadcasted through the UOCS system, thus avoiding the communication error of 5\%.

\section{Simulation of Mephisto observation control}

The Multi-channel Photometric Survey Telescope (Mephisto) has a primary mirror of aperture 1.6 m, and it is equipped with three high-precision charge-coupled device cameras(~\cite{li2020}). It can obtain data amounting to 3.8 GB in a single exposure (nearly 6 TB of data per night) and transmits them to the data center in real time. After online processing through DHS, if a transient source is found, all users are notified via the internet. Additionally, Mephisto can stop observation of the current target and start observation of an emergency target, accessed via internet.

Mephisto is aimed at obtaining the spectral data of billions of stars and galaxies, and millions of quasars through a 10-year survey. To achieve this goal, Mephisto must operate efficiently and obtain the maximum number of observation targets possible within a specified time. Thus, Mephisto has the characteristics of short exposure, high switching frequency of observation targets, high data obtention rate, and a rich telescope communication. Therefore, a stable, real-time, and rapid anomaly analysis and processing of the OCS system are required. 

The communication protocol and interface between the UOCS and various devices, and the observation mode and strategy of the Mephisto are determined through several meetings and discussions. Figures 22 and 23 depict the observation script and control logic of the Mephisto based on UOCS. Differing from LAMOST, every target of Mephisto is exposed twice using the same set of filters. The telescope needs to be slightly adjusted after the first exposure, prior to beginning the second exposure. During the second image readout, the UOCS control telescope points toward the next target, thus maximizing the observation time.

\begin{figure}[!htbp]
	\centering
	\includegraphics[width=6cm, angle=0]{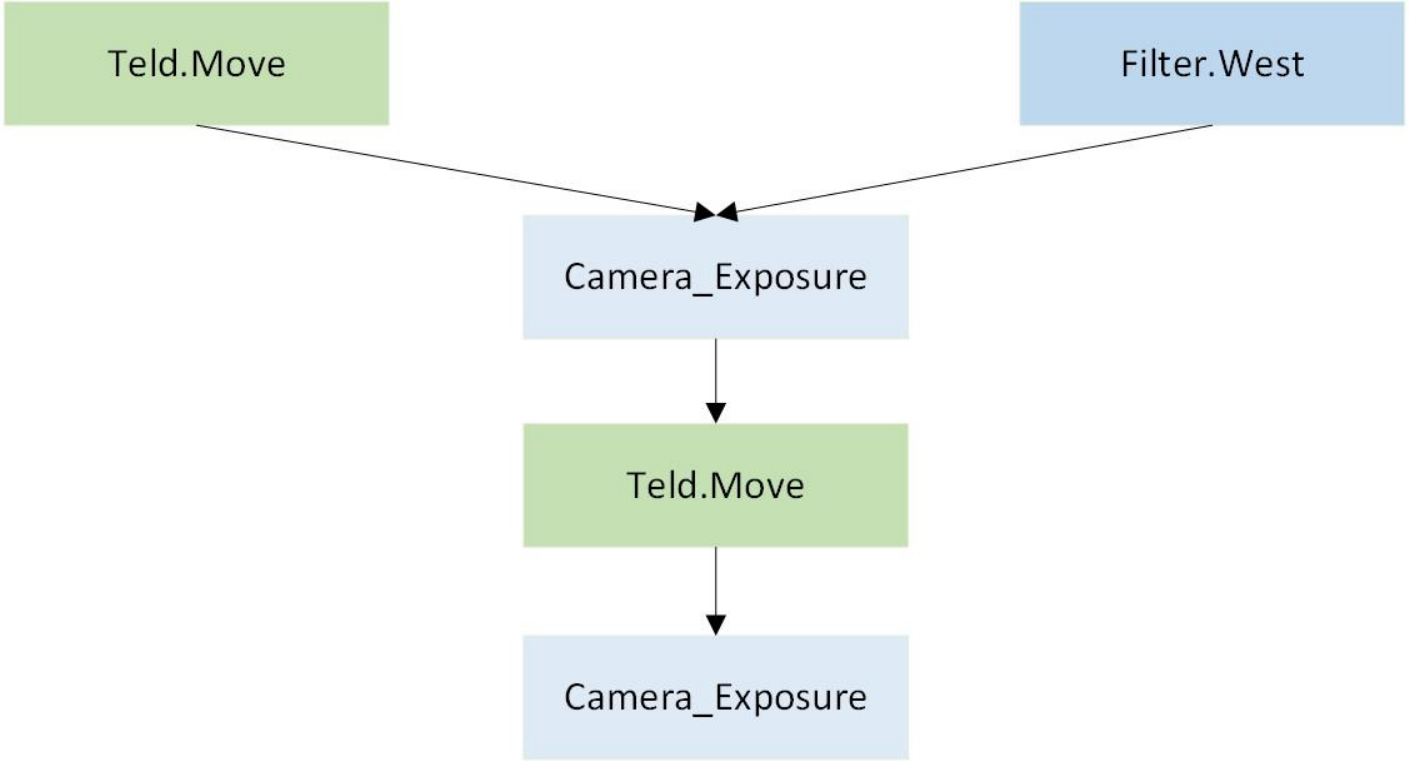}
	\caption{Flow chart of Mephisto observation script}
	\label{Fig22}
\end{figure}

\begin{figure}[!htbp]
	\centering
	\includegraphics[width=12cm, angle=0]{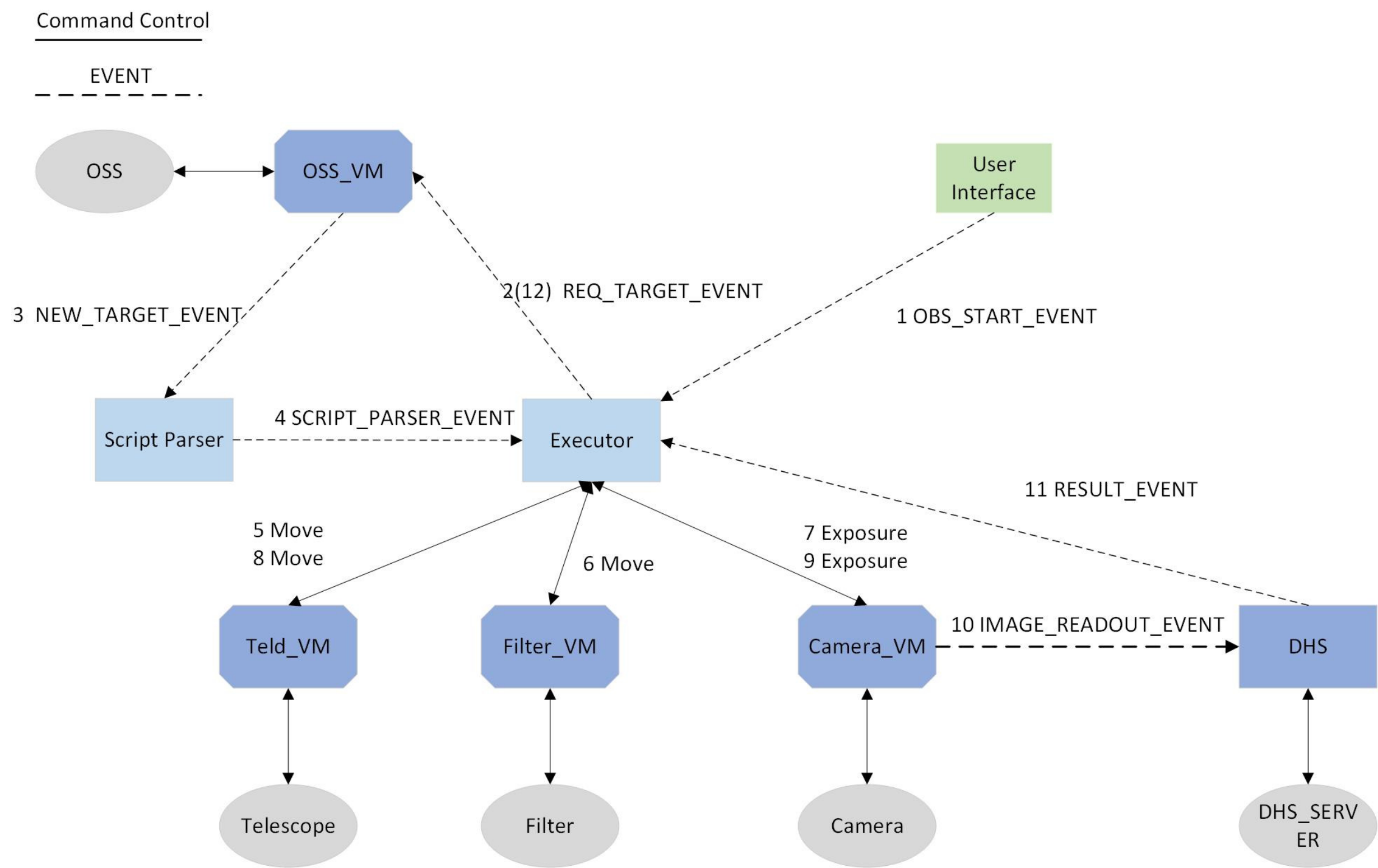}
	\caption{Observation control logic of Mephisto}
	\label{Fig23}
\end{figure}

We built a virtual observation and test environment of Mephisto based on UOCS and began a series of simulation works. Figure 24 shows the execution process of the observation of a target by Mephisto. The green, blue, and gray blocks, respectively, indicate the commands that are completed, being executed, and awaiting execution. To verify the exception handling ability of UOCS, so as to prepare for its later application to Mephisto, we simulated various abnormal scenarios that could arise in various devices of Mephisto, based on the maintenance experience gained from LAMOST.

\begin{figure}[!htbp]
	\centering
	\includegraphics[width=10cm, angle=0]{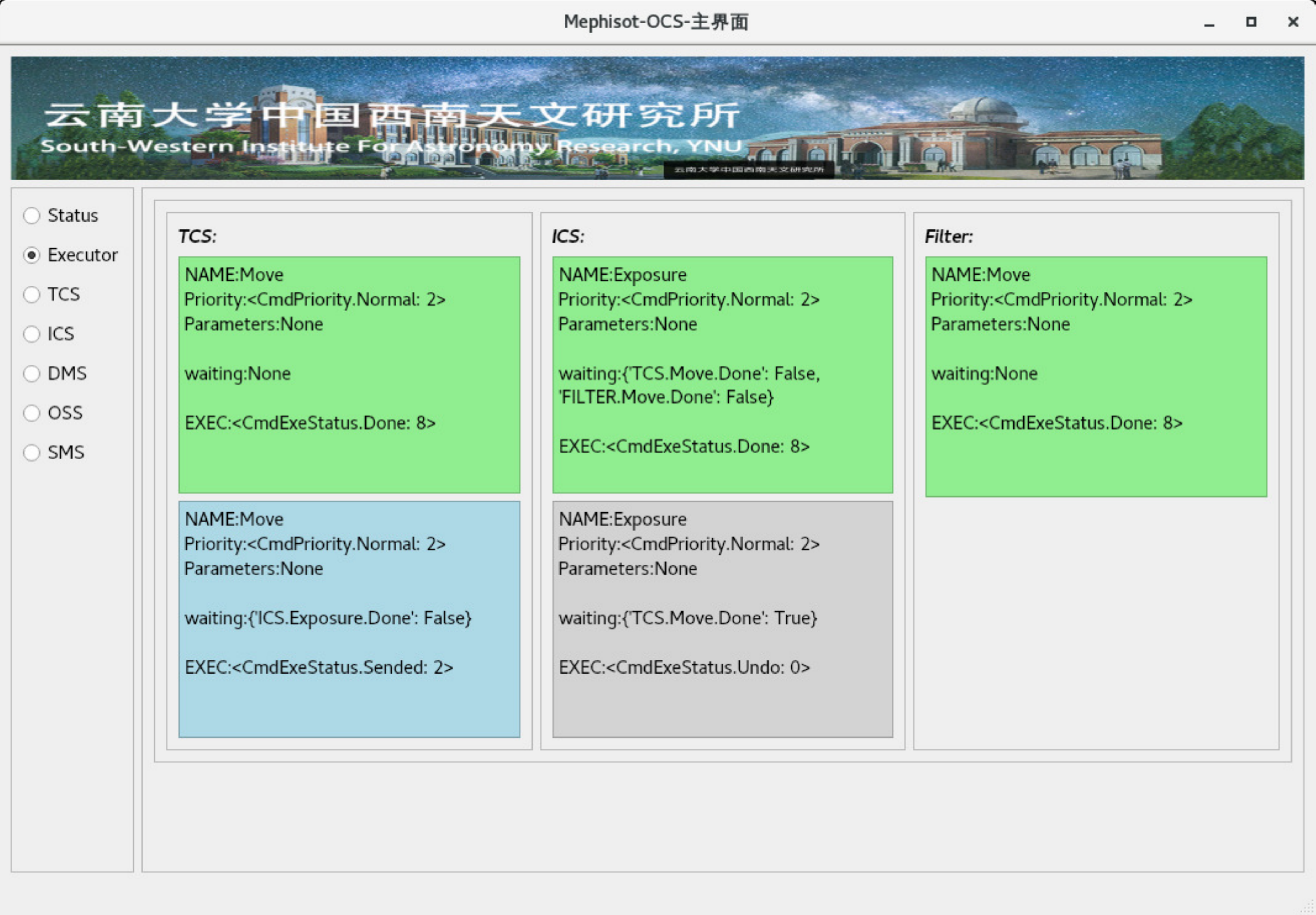}
	\caption{Execution of target observation in Mephisto}
	\label{Fig24}
\end{figure}

\section{Future work}

As for LAMOST, although the unified management of the device running status is realized through UOCS, the communication obstacles between the devices and subsystems are eliminated, and the automatic observation process between the modules is completed preliminarily—this reduces the occurrence of observation accidents caused by communication errors between observers. However, the automatic control of only the camera cluster system and fiber system has been realized as of now, whereas the other devices and subsystems still require to be controlled manually via dialog-box prompts of UOCS. Therefore, longer strides need to be made for a fully automated observation control of LAMOST. In the future, we plan to realize the unified management of all devices and subsystems first, which can be controlled manually by the UOCS to reduce the number of observers and operational steps. Subsequently, we will realize the fully automated control of the subsystems in individual steps and improve the observation efficiency.

The development and construction of each device of Mephisto are progressing steadily. The UOCS will complete offline simulations, start online testing of some devices, and perform a comprehensive system debugging after the completion of telescope construction, ultimately realizing the automated observation and control of Mephisto by 2021.

\section{Conclusion}

The UOCS uses multiple command queues to implement the serial and parallel controls of multiple devices through observation scripts based on device commands. Further, it uses the event-driven messaging mechanism to implement the control logic among the modules in the system. The device command control based on the coroutine can dynamically adjust the command execution according to the priority of the commands. The global state ensures the correct execution of mutually exclusive commands between the devices. The concept of distributed control ensures the stable operation of the entire system to the maximum extent possible.

Currently, UOCS has been implemented for the automated control of the camera cluster system and fiber system of LAMOST. As of now, according to the control logic of Mephisto, the observation process has been simulated. In the future, it will be used for the observation control of Mephisto.

\begin{acknowledgements}
This study is supported by the National Natural Science Foundation of China (Grant Nos. 11703044).  The Guo Shou Jing Telescope ( the Large Sky Area Multi-Object Fiber Spectroscopic Telescope, LAMOST) is a National Major Scientific Project built by the Chinese Academy of Sciences. Funding for the project has been provided by the National Development and Reform Commission. LAMOST is operated and managed by the National Astronomical Observatories, Chinese Academy of Sciences. We also thank the reviewers for suggestions that improved the paper. The authors also gratefully acknowledge the helpful comments and suggestions of the reviewers.
\end{acknowledgements}

%\clearpage

\label{lastpage}

\end{document}